\documentclass[aps,pre,twocolumn]{revtex4-2}
\usepackage{blindtext}
\usepackage{amsmath,amsthm,amsfonts,amssymb,amscd}
\usepackage{verbatim}
\usepackage{mathrsfs}
\usepackage[dvipsnames]{xcolor}
\usepackage{graphicx}
\usepackage{listings}
\usepackage{tikz}
\usepackage{physics}
\usepackage{epigraph}
\usepackage{float}
\usepackage{mathtools}
\usepackage{caption}
\usepackage[caption=false]{subfig}
\usepackage{bm}
\usepackage[capitalise]{cleveref}
\usepackage[export]{adjustbox}
\newcommand{\pd}{\partial}
\newcommand{\ep}{\epsilon}
\newcommand{\vel}{\textbf{v}}
\newcommand{\velt}{\text{v}}

\newcommand{\wrt}{\mathrm{d}}
\usepackage{xcolor} 
\newcommand{\dr}{\wrt \textbf{r}}
\newcommand{\ds}{\wrt \text{S}}
\newcommand{\rp}{\text{Re}}
\newcommand{\imag}{\text{Im}}
\newcommand{\Qt}{\textbf{Q}}

\newcommand{\wb}{\overline{w}}
\newcommand{\zb}{\bar{z}}
\newcommand{\ci}{\mathrm{i}}

\begin{document}
\title{Controlling wall particle interactions with activity}
\author{Luke Neville, Jens Eggers, Tanniemola Liverpool}
\email{luke.neville@bristol.ac.uk}
\affiliation{School of Mathematics, Fry Building, University of Bristol, BS8 1UG, United Kingdom}
\begin{abstract}
    We calculate the effective forces on hard disks near walls embedded inside active nematic liquid crystals. When the disks are sufficiently close to the wall and the flows are sufficiently slow, we can obtain exact expressions for the effective forces.  We find these forces and the dynamics of disks near the wall depend both on the properties of the active nematic and on the anchoring conditions on the disks and the wall. Our results show that the presence of active stresses attract planar anchored disks to walls if the activity is extensile, and repel them if contractile. For normal anchored disks the reverse is true; they are attracted in contractile systems, and repelled in extensile ones.
\end{abstract}
\maketitle

\section{Introduction}
\label{sec:introduction}
Active fluids are a class of complex matter where the individual components consume energy to perform mechanical work~\cite{marchettiHydrodynamicsSoftActive2013, ramaswamy2010mechanics}.  They show a rich phenomenology, from flocking~\cite{Vicsek1995,toner1998flocks,toner2005hydrodynamics}, to non-equilibrium phase transitions and novel emergent collective behaviour~\cite{Fily2012,Cates2015,alert2020universal, alert2022active, creppy2015turbulence,dunkel2013fluid}. To date, most work has focused on understanding this behaviour, however, a major challenge going forward, is how to {\em control} activity, i.e.  which components of a system are active, when that activity is to be switched on/off, how to use it to steer emergent collective behaviour towards a desired function and possibly with a view to even extract useful work~\cite{gompper20202020,maggi2017self}. 
\textbf{}One promising way that has been proposed to do this is using {\em active} colloids, a suspension of microscopic self-driven particles, in a {\em passive} solvent. By controlling activity, the colloids can in principle be designed to self-assemble efficiently ~\cite{bishop2023active,goodrich2017braid,zottl2016emergent} and one can even start to think about how to design soft micro-machines ~\cite{zottl2023modeling,aranson2013active,aubret2021metamachines}.

In this work we are interested in a complementary class of systems, passive particles immersed in an active matrix~\cite{dogiccolloidaps,mallory2014curvature}. To be concrete we focus on active nematics, where the activity is introduced on hydrodynamic scales, in terms of contractile or extensile stresses~\cite{simha2002hydrodynamic,activenematicreview}. These stresses appear naturally in suspensions of active anisotropic particles such as bacterial colonies \cite{dunkel2013fluid,dell2018growing}, the cell cytoskeleton~\cite{FrenchGermanReview07} and in synthetic biomimetic systems such as mixtures of microtubules, polymers and kinesin~\cite{needleman2017active,sanchez2012spontaneous}. Unlike isotropic fluids, nematics have additional topological constraints \cite{mermin1979topological}, which  are perhaps most obvious when particles are added. The anchoring of the nematic to the particles surface endows the particle with non-zero topological charge \cite{starkPhysicsColloidalDispersions2001} which must be counteracted by additional point defects outside the particle. These additional defects can radically change the elastic forces, leading to attraction between particles \cite{poulin1997droplet,poulin1998inverted,muvsevivc2017nematic}. These attractive forces can be incredibly strong, and have been used to create self assembling colloidal crystals \cite{muvsevivc2008self,vskarabot2007two}. Once activity is added to this kind of system however, new phenomena appear, which can be controlled by tuning the particle shape, and the anchoring of the nematic liquid crystal on the particle surface ~\cite{ray2023rectified,houston2023cog,dogiccolloidaps}. For example, colloids that would be stationary in passive liquid crystals can be propelled along by their own topological charge~\cite{loewePassiveJanusParticles2022,yao2022topological}. 

The bulk behaviour of such systems can be studied by looking at infinite systems ignoring the effects of boundaries, however as real 
 experimental systems are always confined it is important to understand the role of boundaries, e.g. interactions between particles and walls. Furthermore these interactions can be the starting point to understand the interactions between particles and hence the collective behaviour of many such particles.  The detailed study of such interactions is the goal of this paper. 
 
 In passive liquid crystals the elastic forces can lead to interactions between embedded particles that can be attractive or repulsive\cite{silvestre2004colloidal}, however it is interesting to explore if activity can change the magnitude or even sign of interaction (i.e from repulsive to attractive or vice-versa). 
 
 To study these forces we use a simple two dimensional system, where the particle is a hard disk, located in the vicinity of a flat wall. Although two dimensions is a simplification, many experimental nematic liquid crystal systems are thin films and so are approximately two dimensional \cite{activenematicreview}. Hence an analysis in two dimensions is a natural starting point for a detailed theoretical study of these systems.  Studying this problem analytically poses a challenge because of the multiple connectivity of the domain \cite{nehari2012conformal}, but can be circumvented using appropriate conformal mapping \cite{koberDictionaryConformalRepresentations1957}. Recently, similar problems were tackled using the director description of the nematic by adapting methods used to study of point vortices in potential flow \cite{chandler2023exact, crowdy2020solving}. However, these approaches are well suited mainly for the large wavelength behaviour and struggle to resolve details of the nematic configurations in the vicinity of the inclusion. We use a complementary approach to describe the nematic, the 2nd rank $\Qt$-tensor, which captures the near-field as well as the far-field behaviour, and in which there is also a natural regularization of topological defects. By explicitly obtaining the $\Qt$ tensor configurations and using them to calculate the forces on the disk, we find that a disk near a flat wall is always repelled by the elastic forces, but may be attracted or repelled by active ones. What controls this depends on a combination of activity and anchoring, and is reminiscent of the anchoring effects in channel flows \cite{batista2015effect,sanchez2012spontaneous,shendruk2017dancing}. 

The rest of the paper is structured as follows, in section \ref{sec:statement} we introduce our model of a particle near a wall. In section \ref{sec:solution} we solve for the nematic director around the particle and use it to calculate the forces, finding that active forces can lead to an attraction or repulsion of the particle. In section \ref{sec: matched analysis} we perform a matched asymptotic analysis, finding that the force is dominated by a small region under the disk. In section \ref{sec: defects} we discuss the location of the topological defects and their effect on the force, before finally discussing our results in section \ref{sec:discuss}.

\section{Model and Statement of Problem}
\label{sec:statement}

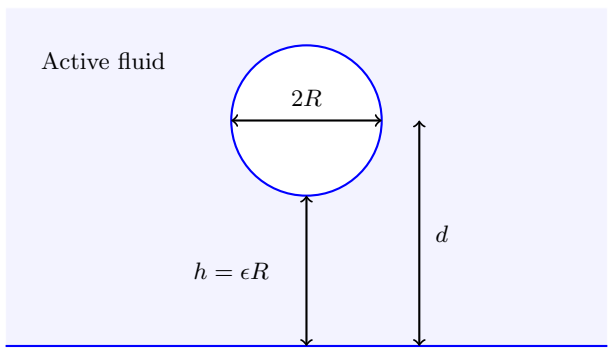
\begin{figure}[h]
\centering
\begin{tikzpicture}
\filldraw[color=blue!5,fill=blue,opacity=0.05] (3,-2) rectangle (11,2.5);
\draw[color=blue,thick] (3,-2)--(11,-2);
\filldraw[color=blue,fill=white, thick](7,1) circle (1);
\draw [to-to,thick](7,-2)--(7,0);
\node at (6,-1){$h=\ep R$};
\draw [to-to,thick](6,1)--(8,1);
\node at (7,1.3){$2R$};
\draw [to-to,thick](8.5,-2)--(8.5,1);
\node at (8.8,-.5){$d$};
\node at (4.3,1.8){Active fluid}; 
\end{tikzpicture}
\caption{A disk sitting in an active fluid above a wall.}
\label{fig:disk wall}
\end{figure}

To study wall-particle interactions we use the set up in \cref{fig:disk wall} consisting of a large circular particle of radius $R$ placed inside in incompressible active nematic at a distance $d$ from a flat uniform wall. The nematic strongly anchors to the wall and surface of the disk, and by symmetry the disk moves in a direction perpendicular to the wall.
%with an as yet unidentified speed.

To model the nematic we use the equations of active nematodynamics describing the coupling of an incompressible velocity field $\vel$ to the elasticity of the nematic order parameter $\Qt$, a symmetric trace-less two tensor \cite{marchettiHydrodynamicsSoftActive2013} in the one Frank elastic constant approximation. The two components of the order parameter $\text{Q}_{xx}=\text{Q}_1$, and $\text{Q}_{xy}=\text{Q}_2$ may be related to the angle of the nematic director using $\theta=\frac{1}{2}\arctan(\text{Q}_2/\text{Q}_1)$, and the scalar order of the nematic by $Q^2_0=\text{Q}_1^2+\text{Q}_2^2$.
\begin{subequations}
\begin{gather}
D_t\Qt=-\frac{1}{\gamma}\frac{\delta \mathcal{F}}{\delta \Qt},\\
\eta\nabla^2\vel-\nabla P+\alpha\nabla\cdot\Qt+\nabla\cdot\boldsymbol{\sigma}^{K}=0.
\end{gather}
\label{eqns: active nematodynamics}
\end{subequations}
The internal structure of the nematic means that we must account for advection as well as rotation of the nematic director. This is given by the first equation, where $\Qt$ relaxes to a minimum of its free energy, $\mathcal{F}$, while being advected and co-rotated by the generalised derivative $D_t$ \cite{beris1994thermodynamics}. The rate of this relaxation is governed by the rotational viscosity $\gamma$. Alongside the equation for $\Qt$ we have the Stokes equations for the fluid, where $P$ is the pressure, and $\eta$ is the viscosity. In addition to the viscous stresses, $\sigma^v_{ij}= \eta (\partial_i u_j + \partial_j u_i) - P \delta_{ij}$ there are additional elastic and active stresses from the nematic. The active stresses $\boldsymbol{\sigma}^\alpha=\alpha\Qt$ come from the local flows around each nematogen, with $\alpha < 0 (>0)$ for locally contractile (extensile) flows. The elastic stresses $\boldsymbol{\sigma}^K$ derive from the Landau- de Gennes free energy
\begin{equation}
\mathcal{F}=\int\dr\ \frac{K}{2}(\nabla_i\text{Q}_{jk})(\nabla_i\text{Q}_{jk})-\frac{A}{2}\text{Q}_{ij}\text{Q}_{ij}+\frac{A}{4}(\text{Q}_{ij}\text{Q}_{ij})^2,
\label{eqn: Landau de Gennes}
\end{equation}
where $K$ is the elastic constant of the nematic, and $A$ controls the strength of the nematic order \cite{gennesPhysicsLiquidCrystals2013}.

%The equation for $\Qt$ is highly non-linear and therefore difficult to study analytically. 
We make two simplifying approximations, the first being that the radius of the disk is smaller than the elastic screening length $L_K^2=A/K$. Assuming this we may neglect the $A$ terms in the free energy and take the harmonic approximation \cite{thampi2015intrinsic}
\begin{equation}
\mathcal{F}=\int\dr\ \frac{K}{2}(\nabla_i\text{Q}_{jk})(\nabla_i\text{Q}_{jk}).
\end{equation}
Importantly this approximation is only valid when there is a wall, and the harmonic approximation to $\Qt$ does not work well for an isolated disk.

Secondly, we assume the advective time scale to be much longer than the diffusive time scale. By comparing advective and diffusive terms in \cref{eqns: active nematodynamics} we find this is true when $h^2\ll K\eta/\alpha\gamma$, where $h$ is the closest distance between the disk and wall. This is the same condition for no spontaneous active flows in a channel of thickness $h$ \cite{voituriez2005spontaneous}. Although the region under the disk is curved we expect similar qualitative behaviour and so all active flows under the disk ought to be weak.

Assuming these two conditions leads to a simpler, linear set of equations \cite{loisy2019tractionless},
\begin{subequations}
\begin{gather}
\nabla^2\Qt=0\\
\eta\nabla^2\vel-\nabla P+\alpha\cdot\nabla\Qt+\nabla\cdot\boldsymbol{\sigma}^{K}=0,\\
\sigma_{i j}^{K}=-K\left[\nabla_i \text{Q}_{k l} \nabla_j \text{Q}_{k l}-\frac{1}{2} \delta_{i j}\left|\nabla_j \text{Q}_{k l}\right|^2\right]
+[\Qt,\textbf{H}]_{ij},
\end{gather}
\label{eqn:q_eqns_simple}
\end{subequations}
where $\textbf{H}=\delta \mathcal{F}/\delta\Qt$, and the elastic stress is derived from the harmonic free energy in the Appendix. 

As the speed of the disk is as yet unknown, the velocity boundary conditions are zero velocity on the wall, and $\vel=\textbf{U}$ on the disk. $\textbf{U}$ will later be found with the condition that the disk be force free \cite{laugaHydrodynamicsSwimmingMicroorganisms2009}, $\text{F}_i = \displaystyle \int_{\mbox{\tiny Disk}} \left( \sigma_{ij}^v +\sigma_{ij}^a +\sigma_{ij}^K\right) \text{n}_j \ds=0$. The boundary conditions for $\Qt$ are determined by the anchoring on the wall and disk, which we assume to be strongly planar or normal. From now on we shall work in units where $R=1$ unless explicitly stated.

To calculate the hydrodynamic forces on the disk we will first solve for $\Qt$, then substitute it into the Stokes equations to find the stresses.

\section{Solution}
\label{sec:solution}

\subsection{Nematic field}
\label{sec: Q tensor sol}
We assume that the nematic strongly anchors to all surfaces, giving boundary conditions
\begin{subequations}
\begin{gather}
\text{Q}_{1}=\pm \cos 2\theta_d\ \text{on the disk},\\
\text{Q}_{2}=\pm \sin 2\theta_d\ \text{on the disk},\\
\text{Q}_{1}=\mp 1\ \text{at the wall and infinity},\\
\text{Q}_{2}=0\ \text{at the wall and infinity},
\end{gather}
\label{eqn:Q boundary conditions}
\end{subequations}
where $\theta_d$ is the polar angle centred on the disk. The $+$ gives normal anchoring, and the $-$ planar. Comparing the $\Qt$ boundary conditions and the definition of the active stress, we see that switching from planar to normal anchoring is equivalent to switching the sign of $\alpha$. In what follows we shall assume planar anchoring for all calculations.

\begin{figure}[h]
\centering
\begin{tikzpicture}
\filldraw[color=blue,fill=blue!5, thick](7,1) circle (2.5);
\filldraw[color=blue,fill=white, thick](7,1) circle (1);
\draw [to-to,thick](6,1)--(8,1);
\draw [to-to,thick](10,-1.5)--(10,3.5);
\node at (10.5,1){$2$};
\node at (7,1.3){$2\rho$};
\node at (7,2.75){$w$ plane};
\filldraw[color=blue,fill=blue, thick](4.5,1) circle (.1);
\node at (3.8,1){$w=-1$};
\end{tikzpicture}
\caption{Conformally mapped domain. The inner circle corresponds to the disk boundary, the outer circle to the wall, and $w=-1$ to the point at infinity.}
\label{fig:conformal map}
\end{figure}
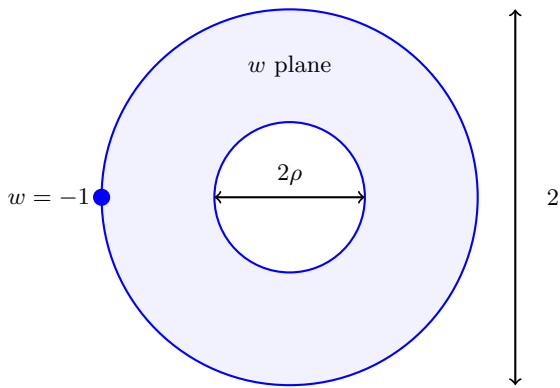

To solve for $\Qt$ we use conformal mapping, noting that Laplace's equation is invariant under any conformal map \cite{nehari2012conformal}. Defining $z=x+\ci y$, the relevant conformal mapping is \cite{crowdyTreadmillingSwimmersNoslip2011}
\begin{equation}
w(z)=\frac{s+\ci z}{s-\ci z} \ \Leftrightarrow\ z(w)=\ci s\frac{1-w}{1+w},
\label{eqn:conformal map}
\end{equation}
where $s$ depends on $d=1+h$ through
\begin{subequations}
\begin{gather}
s=\frac{1-\rho^2}{2\rho},\\
\rho=d-\sqrt{d^2-1}.
\end{gather}
\end{subequations}
This map takes the blue region in \cref{fig:disk wall} to an annulus of inner radius $\rho$ and outer radius $1$, see \cref{fig:conformal map}. The wall maps to $|w|=1$, the disk to $|w|=\rho$, and the point at infinity to $w=-1 $.

\begin{figure}[h]
    \centering
    \subfloat{
  \includegraphics[clip,height=5.5cm]{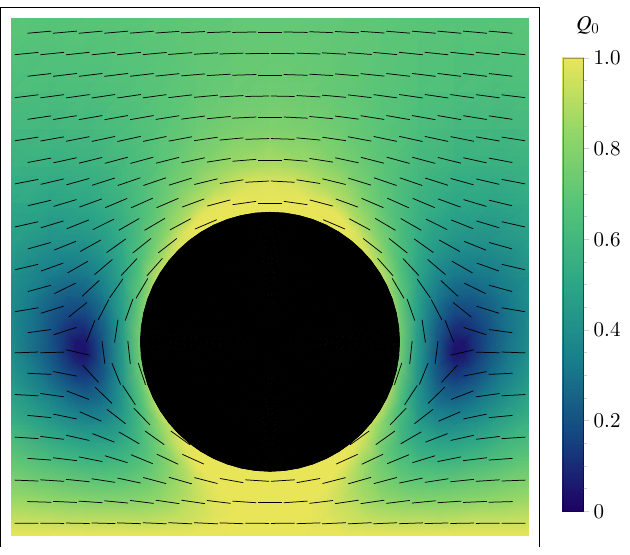}%
}

    \subfloat{
  \includegraphics[clip,height=5.5cm]{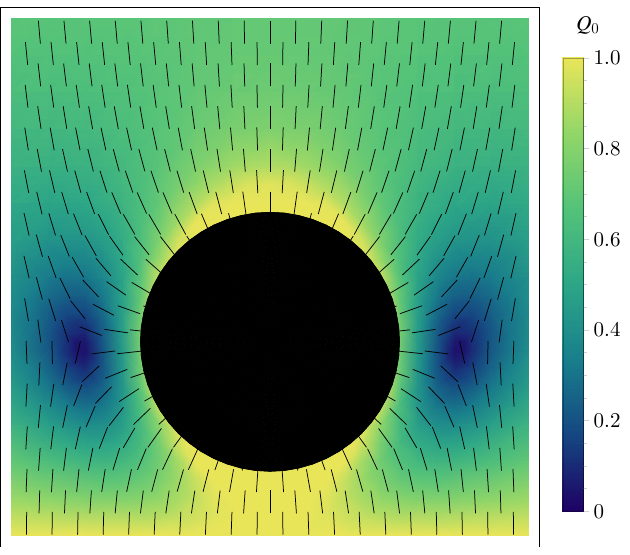}%
}
    \caption{Nematic texture around the disk when $d=1.5R$. (Top) Planar anchoring, (Bottom) Normal anchoring.}
    \label{fig:Q tensor}
\end{figure}

Defining the complex field, $Q=\text{Q}_{1}+\ci\text{Q}_{2}$, the solution is found using polar coordinates in the annulus, and is equivalent to a Laurent series in $w$ and $\wb$,
\begin{equation}
Q=\left(1+\frac{(\rho-1)\log w\wb}{2\log\rho}+\sum_{n=1}^{\infty}c_n\left[w^n-\frac{1}{\overline{w}^n}\right]\right),
\label{eqn:q soln}
\end{equation}
where the coefficients $c_n$ are
\begin{equation}
c_n=\frac{(-1)^n}{1-1/\rho^{2n}}[(1+n)\rho^2-2n+(n-1)\rho^{-2}].
\label{eqn: cn coeffs}
\end{equation}
$\Qt$ in real space is then given by $Q(w(z))$. Examining plots of this solution in \cref{fig:Q tensor} we see two $-1/2$ topological defects slightly below the equator of the disk, with their position given by the zeros of $Q_0$ \cite{chaikin1995principles}. These are needed to counteract the net $+1$ topological charge of the disk and represent the lowest energy configuration of defects. The series solution in (\ref{eqn:q soln}) may be interpreted as a multipole expansion, with the  n\textsuperscript{th} term corresponding to the n\textsuperscript{th} multipole \cite{houston2023multi}. For large disk-wall separations the solution converges rapidly as only the first two terms are needed to capture the quadrupole defect structure.

\subsection{Reciprocal theorem}
With $\Qt$ found we use it to calculate the forces on the disk. To do this we will not need to calculate the velocity field using \cref{eqn:q_eqns_simple}, but shall instead apply the Lorentz reciprocal theorem \cite{masoudReciprocalTheoremFluid2019, kimMicrohydrodynamicsPrinciplesSelected2005}. The Lorentz reciprocal theorem relates the stress tensors in two solutions of the Stokes equations in the same domain but with different boundary conditions. To derive the result, consider two copies of the Stokes equations in a domain $\mathcal{D}$
\begin{subequations}
    \begin{gather}
        \eta\nabla^2\vel-\nabla P+\nabla\cdot\boldsymbol{\Sigma}=0,\\
        \eta\nabla^2\hat{\vel}-\nabla\hat{P}=0,
    \end{gather}
\end{subequations}
where only one of the copies contains extra stresses $\boldsymbol{\Sigma}$, and we have used a carat to distinguish the secondary problem variables. Using these and incompressibility the following integral identity holds
\begin{equation}
    \int_{\mathcal{D}}\nabla_i[\hat{\velt}_j\sigma_{ij}]-\nabla_i[\velt_j\hat{\sigma}_{ij}]\dr = \int_{\mathcal{D}}\Sigma_{ij}\nabla_i\velt_j\wrt r,
\end{equation}
where we have defined the two stress tensors
\begin{subequations}
\begin{gather}
\sigma_{ij}=\eta[\nabla_i\text{u}_j+\nabla_j\text{u}_i]-P\delta_{ij}+\Sigma_{ij},\\
\hat{\sigma}_{ij}=\eta[\nabla_i\hat{\text{u}}_j+\nabla_j\hat{\text{u}}_i]-\hat{P}\delta_{ij}.
\end{gather}
\end{subequations}
Applying the divergence theorem to the above integral yields the modified Lorentz reciprocal theorem
\begin{equation}
\int_{\pd \mathcal{D}}[\hat{\text{u}}_i\sigma_{ij} -\text{u}_i\hat{\sigma}_{ij}]\text{n}_j\ds= -\int_{\mathcal{D}} \Sigma_{ij}\nabla_i\hat{\text{u}}_j\dr,
\label{eqn: recip theorem unintegrated}
\end{equation}
where $\pd\mathcal{D}$ is the boundary of $\mathcal{D}$, $\textbf{n}$ is the unit normal to the boundary directed into the fluid domain, and $\ds$ is the boundary surface element.

To apply this theorem we use the boundary conditions for $\vel$, and choose boundary conditions for $\hat{\vel}$ to be zero velocity on the wall, and $\hat{\vel}=\hat{\textbf{U}}$ on the disk. Substituting these into \cref{eqn: recip theorem unintegrated} we find
\begin{equation}
    \hat{\textbf{U}}\cdot\textbf{F}-\textbf{U}\cdot\hat{\textbf{F}}=-\int_{\mathcal{D}} \Sigma_{ij}\nabla_i\hat{\text{u}}_j\dr,
\end{equation}
where we identified the total (non-viscous) force on the disk $\textbf{F}$ as the integral of the (non-viscous) stress on its surface \cite{batchelorIntroductionFluidDynamics2010}. From the (translational) symmetry of the problem we expect the force to be perpendicular to the wall, so choosing $\hat{\textbf{U}}=\textbf{e}_y$ we have
\begin{equation}
    \textbf{F}_y = \textbf{U}\cdot\hat{\textbf{F}}-\int_{\mathcal{D}} \Sigma_{ij}\nabla_i\hat{\text{u}}_j\dr.
\label{eqn: reciprocal theorem forces}
\end{equation}
This form of the reciprocal theorem is most helpful to us. 

To apply (\ref{eqn: reciprocal theorem forces}) to the active nematic problem we set $\boldsymbol{\Sigma}=\alpha\Qt+\boldsymbol{\sigma}^K$ (non-viscous part of the stress) and then solve for the secondary velocity $\hat{\vel}$ using
\begin{subequations}
    \begin{gather}
        \eta\nabla^2\hat{\vel}-\nabla\hat{P}=0,\\
        \hat{\vel}=\textbf{0}\ \text{on the wall},\\
        \hat{\vel}=\textbf{e}_y\ \text{on the disk}.
    \end{gather}
\label{eqn: secondary velocity equations}
\end{subequations}
We are able to take advantage of the fact that the solution to equation (\ref{eqn: secondary velocity equations}) is already known, having first been found by Jeffrey and Onishi \cite{jeffreySLOWMOTIONCYLINDER1981}, and later re-derived by Crowdy using complex analysis \cite{crowdyTreadmillingSwimmersNoslip2011}. The rest of the paper is devoted to performing the integrals in \cref{eqn: reciprocal theorem forces} a task which is the main technical result of this paper.

\subsection{Active force}
The linearity of \cref{eqn: reciprocal theorem forces} allows the (non-viscous) force to be split into an active and elastic contribution $\textbf{F}=\textbf{F}^{\alpha}+\textbf{F}^K$, where the active contribution is
\begin{equation}
    \text{F}^{\alpha}=-\alpha\int_{\mathcal{D}}\text{Q}_{ij}\nabla_i\hat{\velt}_j\dr.
\label{eqn: active integral}
\end{equation}
Before performing this integral it is instructive to integrate by parts
\begin{equation}
    \text{F}^{\alpha} = \alpha \hat{\text{U}}_j\int_{\text{Disk}} \text{Q}_{ij}\text{n}_i\ds + \alpha\int_{\mathcal{D}}\hat{\velt}_j\nabla_i \text{Q}_{ij}\dr,
\end{equation}
where we have used the boundary conditions on the disk to factor out the constant $\hat{\textbf{U}}$ in the surface integral. The first term on the right hand side is the
integral of the active stress over the disk, and so contains no information about stress due to active flows. These are
contained in the second term which couples the secondary flow to the driving term in the Stokes equation. It is interesting to contrast with earlier work on particles in active nematics using directors rather Q-tensors. Only the equivalent of the first term was calculated there, as the required secondary solution to the Stokes equations was unknown \cite{houston2023cog, houston2023multi}.

To perform the integral we change to complex coordinates, and integrate in the conformally mapped domain
\begin{equation}
\text{F}^\alpha_y=-\alpha \int_{\mathcal{D}} \text{Q}_{ij}\nabla_i\hat{\text{v}}_j\dr=-2\alpha\int_{\mathcal{D}} \rp\left[\frac{\pd \hat{v}}{\pd \bar{z}} \overline{Q}\right]\dr,
\end{equation}
where $\hat{v}=\hat{\velt}_x+\ci \hat{\velt}_y$ and $\overline{Q}=\text{Q}_1-\ci \text{Q}_2$ is the complex conjugate of $Q$. Doing the integral yields an active force (with dimensions reinstated)
\begin{equation}
\begin{aligned}
\text{F}^\alpha_y&=-\frac{\pi\alpha R (-1+\rho)^2(1+\rho)\left(-1+\rho^4-4\rho^2 \log \rho \right)}{2 \rho^3 \log \rho\left(1-\rho^2+\left(1+\rho^2\right) \log \rho\right)},\\
&\sim 4\sqrt{2}\pi R\alpha \ep^{1/2} +O(\ep).
\end{aligned}
\label{eqn: active force exact}
\end{equation}
%where we have reinstated dimensions $R$. 
These results imply that at short distances, planar anchored disks are repelled by active forces in contractile nematics, and attracted in extensile ones. By switching the sign of $\alpha$ we see the reverse is true in normal anchored nematics.

\subsection{Elastic force}
The elastic contribution to \cref{eqn: reciprocal theorem forces} is
\begin{equation}
\begin{aligned}
    \text{F}^K&=-\int_{\mathcal{D}} \sigma^K_{ij}\nabla_i\hat{\velt}_j\dr\\
    &=\int_{\text{Disk}}\sigma^K_{ij}\hat{\velt}_i\text{n}_i\ds+\int_{\mathcal{D}}\hat{\velt}_j\nabla_i\sigma^K_{ij}\dr.
\end{aligned}
\end{equation}
where we have integrated by parts. Using $\nabla^2\Qt=0$ and the definition of $\boldsymbol{\sigma}^K$, the second term vanishes. The elastic forces therefore do not affect the flow and are given by the surface integral of $\boldsymbol{\sigma}^K$ over the disk
\begin{equation}
\text{F}^K=\hat{\text{U}}_j\int_{\text{Disk}}\sigma^K_{ij} \text{n}_i\ds.
\label{eqn: elastic integral}
\end{equation}
To perform the integral we write the stress and surface normal vectors as complex variables $\sigma^K=\sigma_{11}+\ci\sigma^K_{12}$ and $\text{n}=\text{n}_1+\ci\text{n}_2$, which after substitution give
\begin{align}
\text{F}^K&=-4K\oint_{\text{Disk}}\imag[\sigma\bar{\text{n}}]\ds,\\
&=-4K\oint_{\text{Disk}}\rp\left[\frac{\pd Q}{\pd \bar{z}}\frac{\pd \overline{Q}}{\pd\bar{z}}\frac{\wrt \bar{z}}{\ds}\ds \right].
\label{eqn:Q elastic integral}
\end{align}
We calculate this integral using the conformally mapped coordinates, giving
\begin{equation}
\begin{aligned}
    \text{F}^K=&\frac{8\pi K}{R(1-\rho^2)}\Bigg[2c_0^2+2c_0c_1+\sum_{n=1}^{\infty}2n^2 c_n^2
\\
&+n(n+1)c_nc_{n+1}+n(n-1)c_nc_{n-1}\Bigg],
\end{aligned}
\label{eqn:Q elastic series}
\end{equation}
where $c_0=(\rho-1)/2\log\rho$, and $c_n$ are as defined in (\ref{eqn: cn coeffs}).

As shown in Fig. \ref{fig: elastic force and steady height}(a) the series converges quickly and only two terms in the sum are required for an accurate result; again because two terms are the minimum to capture the quadrupole structure of the defects. The asymptotics for small $\ep$ are also dominated by these terms, with the force diverging as $\ep^{-1/2}$ as the disk approaches the wall due to the increasing curvature in the nematic director.
\begin{equation}
\text{F}^K\sim \frac{2\sqrt{2}K\pi}{R}\ep^{-1/2} +O(1),\\
\label{eqn: elastic force leading order}
\end{equation}

\subsection{Steady state}
\begin{figure}[t]
\centering
  \subfloat{\includegraphics[width=.88\linewidth]{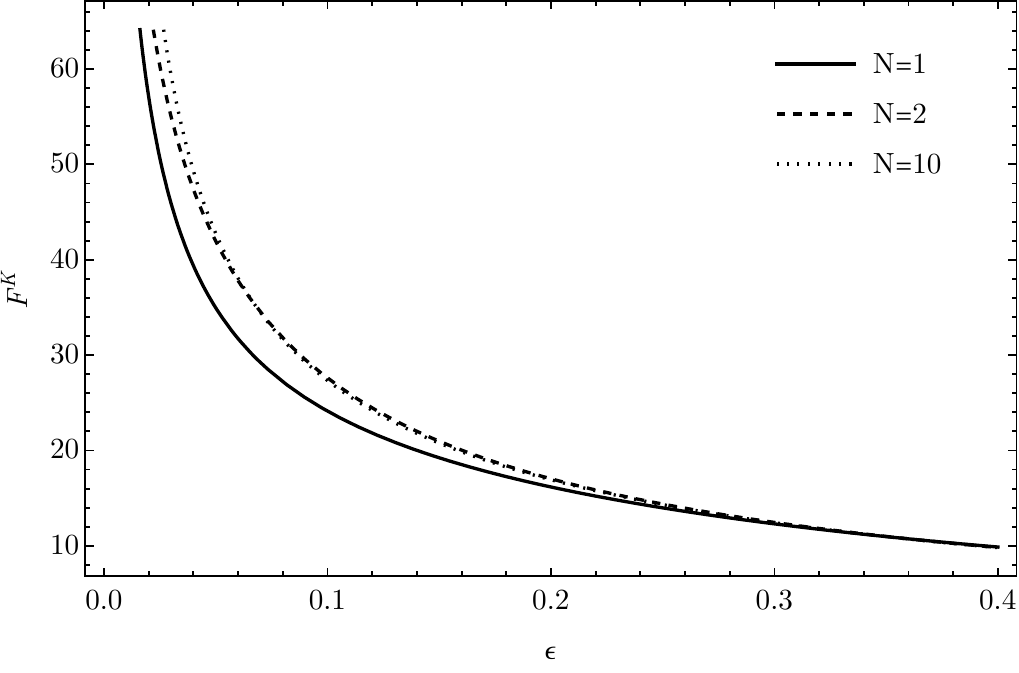}}
  
\subfloat{\includegraphics[width=.88\linewidth]{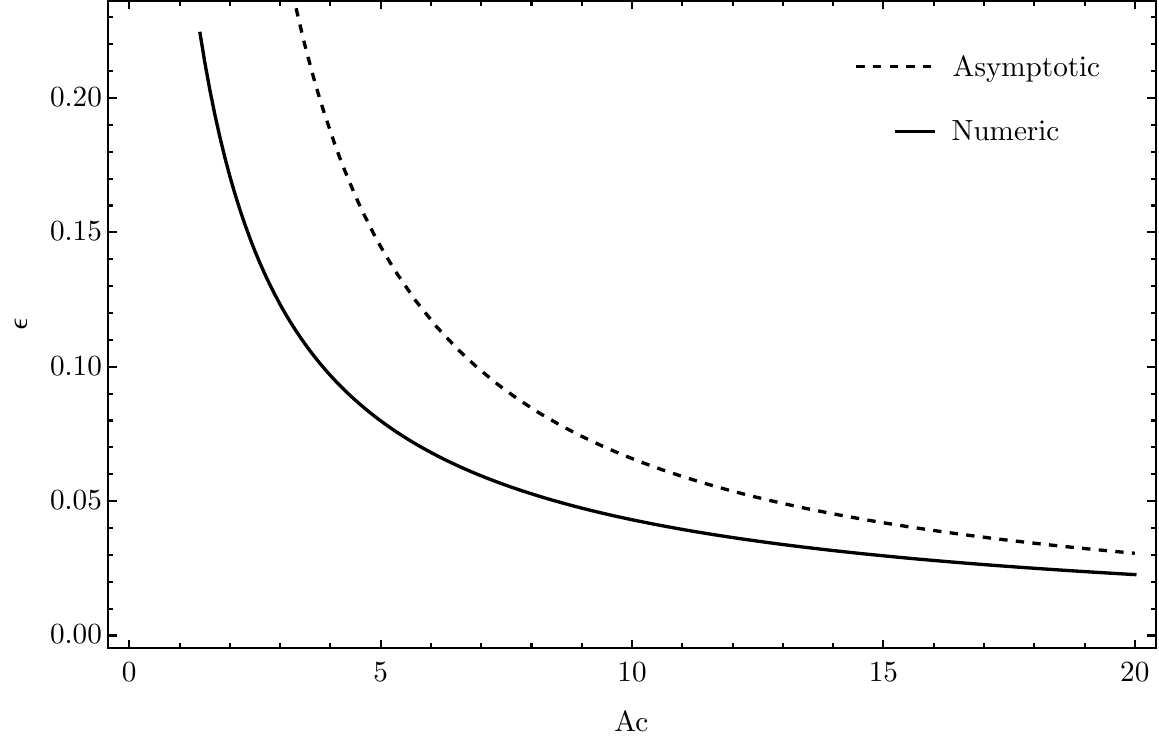}}
\caption{(Top) The elastic force against $\ep$ after including $N$ terms in the summation. After two terms there is an adequate result. (Bottom) The steady state value of $\ep$ against $\text{Ac}=\alpha R^2/K$. The numerical result was calculated using $10$ terms in the series for the elastic force, and the asymptotic value is given by $\ep=1/2\text{Ac}$.}
\label{fig: elastic force and steady height}
\end{figure}

To calculate the steady state separation we substitute the active and elastic forces into Eq. (\ref{eqn: reciprocal theorem forces}), with the requirement that the disk be force free \cite{laugaHydrodynamicsSwimmingMicroorganisms2009}. In steady state, the velocity of the disk, $\textbf{U}$, vanishes, giving the condition $\text{F}^K+\text{F}^{\alpha}=0$. Using this and the sign of the active force, we conclude that planar (normal) anchored disks have a stable steady state height when the nematic is extensile (contractile) as otherwise all forces on the disk are repulsive. Assuming this to be true we numerically solve the force balance condition for $\ep$, with the results shown in Fig. \ref{fig: elastic force and steady height}. These numerical solutions are compared against the asymptotic result  $\ep_{\text{steady}}=K/2\alpha R^2$, derived by balancing the leading order active and elastic terms.

\section{Matched Asymptotic analysis}
\label{sec: matched analysis}
The exact results we have derived give the force on the disk for arbitrary $\ep$, but do not indicate where the force comes from. Is the force dominated by the region under the disk, outside, or is it from both? To investigate this we perform a matched asymptotic analysis of the force integrals (\ref{eqn: active integral}, \ref{eqn: elastic integral}) \cite{cooley1969slow,bender2013advanced, Hinch_1991}.

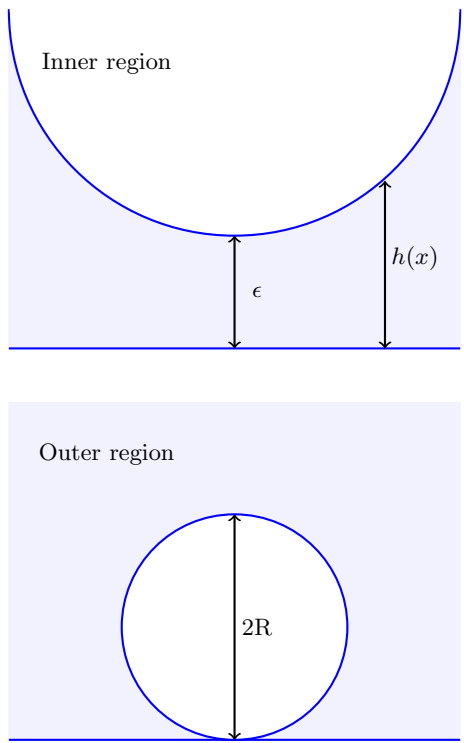
\begin{figure}[h]
\centering
\subfloat{
    \begin{tikzpicture}
    \filldraw[color=blue!5,fill=blue!5] (-3,-2) rectangle (3,2.5);
    \draw[color=blue,thick, fill=white] (-3,2.5) -- (3,2.5) arc(0:-180:3) --cycle;
    \draw[color=white,ultra thick] (-2.99,2.5) -- (2.99,2.5);
    \draw[color=blue,thick] (-3,-2)--(3,-2);
    \draw[to-to, thick] (0,-2) -- (0,-0.5);
    \draw[to-to, thick] (2,-2) -- (2,0.23);
    \node at (2.4,-0.8){$h(x)$};
    \node at (0.3,-1.25){$\ep$};
    \node at (-1.7,1.8){Inner region}; 
    \end{tikzpicture}}

\subfloat{
\begin{tikzpicture}
\filldraw[color=blue!5,fill=blue!5] (4,-2) rectangle (10,2.5);
\draw[color=blue,thick] (4,-2)--(10,-2);
\filldraw[color=white,fill=white] (4,2.5) rectangle (10,2.8);
\filldraw[color=blue,fill=white, thick](7,-0.5) circle (1.5);
\node at (5.3,1.8){Outer region}; 
\draw[to-to, thick] (7,-2) -- (7,1);
\node at (7.3,-0.5){2R};
\end{tikzpicture}}

\caption{Inner and outer regions for matched asymptotic expansion.}
\label{fig: inner and outer regions}
\end{figure}

\subsection{Lubrication analysis}
Before moving onto a formal asymptotic analysis of the problem we first apply a naive lubrication analysis of \cref{eqn:q_eqns_simple}, by which we mean neglecting horizontal derivatives for vertical ones \cite{joanny2012drop}
\begin{subequations}
    \begin{gather}
        \pd_y^2\text{Q}_2=0,\\
        \eta\pd_y^2\velt_x -\pd_x P+\alpha\pd_y \text{Q}_2=0.
    \end{gather}
\end{subequations}
The boundary conditions are $\velt_x=0$ on the disk and wall; $\text{Q}_2=0$ on the wall; and $\text{Q}_2\approx 2\theta_d$ on the wall, where we have expanded the boundary conditions for small disk angle. The distance between the disk and the wall is approximately $h(x)=\ep+x^2/2$, and the disk angle is $h'(x)=x$. Using these we find $\text{Q}_2=2yx/h$, and a horizontal velocity
\begin{equation}
    \velt_x=\frac{P'}{2\eta}y(y-h)-\frac{\alpha x}{h\eta}y(y-h),
\end{equation}
where $P'=\pd_x P$. If the disk moves upwards at velocity $\text{U}$ then flux arguments give the pressure as
\begin{equation}
P=2\alpha\log h-\frac{6\text{U}\eta}{h^2}.
\end{equation}
Unfortunately the active contribution diverges logarithmically. This means that this lubrication theory must be matched to an outer region to get a finite answer. This logarithmic divergence was also found in the study of active nematic droplets, however it did not cause problems because of the finite drop size \cite{joanny2012drop}. Substituting this pressure into the velocity equation, we find all active flow terms cancel out, meaning that the dominant active effect is from an active pressure rather than active flows. This justifies our earlier assumption of neglecting advection at leading order. Interestingly, integrating the pressure under the disk, we find the active force active to be
\begin{equation}
\begin{aligned}
    \text{F}^\alpha &\sim4\pi\sqrt{2}R\alpha\ep^{1/2}+\text{terms that need to be matched}
\end{aligned}
\label{eqn: naive lubrication force}
\end{equation}
i.e. the finite part of the active force gives the correct leading order result.

One may wonder why we referred to the lubrication analysis above as `naive'. This is because a more detailed asymptotic analysis reveals that all $\Qt$ terms are irrelevant to leading order in $\ep$. This more detailed asymptotic analysis is given next.

\begin{comment}
    \begin{figure}[h]
    \centering
    \includegraphics[width=0.5\textwidth]{figures/stresses.pdf}
  \caption{Contractile stresses under the disk. The black bars give schematic positions of individual nematogens, and the black arrows show the contractile flows exerted by each nematogen.}
  \label{fig: active stresses under the disk}
\end{figure}
\end{comment}

\subsection{Inner expansion}
\begin{figure}[t]
\centering
\subfloat{
  \includegraphics[width=.7\linewidth]{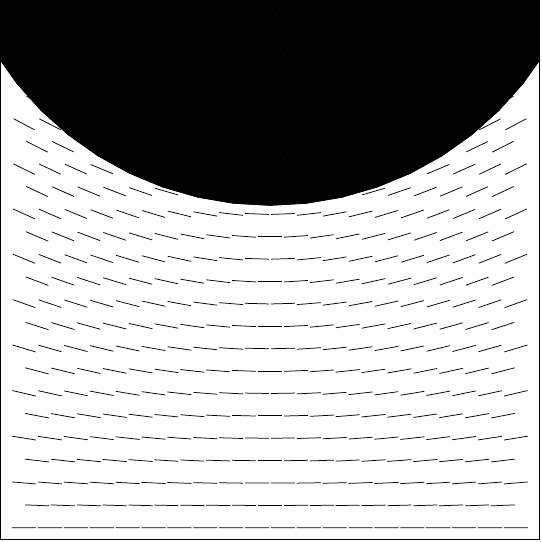}}

\subfloat{
  \includegraphics[width=.7\linewidth]{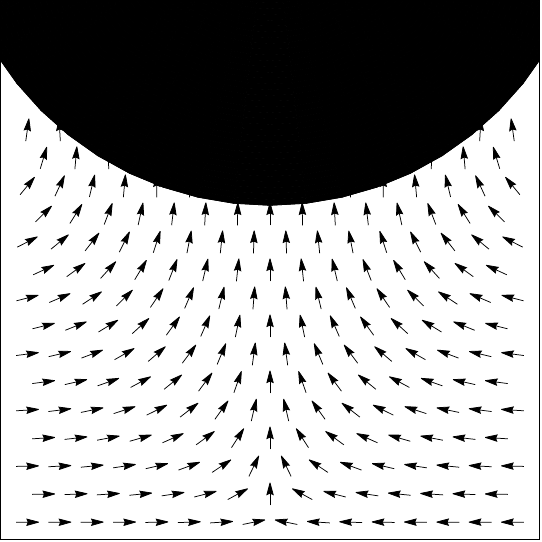}}
\caption{Nematic director and secondary velocity in the inner region. (Top) The nematic director is parallel anchored to all surfaces with $Q_0\approx 1$ everywhere. (Bottom) The velocity is zero on the wall, vertically upwards on the disk, and has a stagnation point at the origin.}
\label{fig: inner velocity and Q}
\end{figure}

The inner region corresponds to a small area under the disk as in \cref{fig: inner and outer regions}. The surface of the disk is at $y=h(x)$, where 
\begin{equation}
    h(x)=1+\ep-\sqrt{1-x^2}\approx \ep+\frac{1}{2}x^2+\frac{1}{8}x^4+o(x^4).
\end{equation}
From this expansion we see that the inner variables are defined by $y=\ep Y$, and $x=\ep^{1/2}X$, in which
\begin{equation}
    \frac{h(x)}{\ep}=H(X)=1+\frac{1}{2}X^2+O(\ep),
\end{equation}
and the angle of the disk is $\theta_d=h'(x)\sim \ep^{1/2}H'(X)$. Substituting these scaled variables into \cref{eqn:q_eqns_simple,eqn: secondary velocity equations,eqn:Q boundary conditions} leads us to pose the expansions
\begin{subequations}
    \begin{gather}
         Q_1=q_{1,0}+\ep q_{1,1}+...,\\
         Q_2=\ep^{1/2}q_{2,0}+\ep^{3/2} q_{2,1}+...,\\
         \hat{\text{v}}_x=\ep^{-1/2}\hat{u}_0+\ep^{1/2} \hat{u}_1+...,\\
         \hat{\text{v}}_y=\hat{v}_0+\ep \hat{v}_1+...,\\
         \hat{P}=\ep^{-3/2}\hat{P}_0+\ep^{-1/2} \hat{P}_1+...,
    \end{gather}
    \label{eqn: inner expansions}
\end{subequations}
where the scaling of $\hat{\velt}_x$ compared to $\hat{\velt}_y$ is from incompressibility. Substitution of these back into \cref{eqn:q_eqns_simple,eqn: secondary velocity equations} yields the leading order equations
\begin{subequations}
    \begin{gather}
        \eta\pd_Y^2 \hat{u}_0-\pd_X \hat{P}_0=0,\\
        \pd_Y \hat{P}_0=0,\\
        \pd_Y^2 q_{1,0}=0,\\
        \pd_Y^2 q_{2,0}=0.
    \end{gather}
\end{subequations}
Solving these with appropriate boundary conditions (\ref{eqn:Q boundary conditions},\ref{eqn: secondary velocity equations}) gives
\begin{subequations}
    \begin{gather}
        q_{1,0}= 1,\\
        q_{2,0}= \frac{2XY}{H},\\
        \hat{u}_0=\frac{6XY(Y-H)}{H^3}\\
        \hat{v}_0=\frac{Y^2\left(3 H^2+6 X^2 Y-2 H\left(3 X^2+Y\right)\right)}{H^4},
    \end{gather}
\label{eqn: inner Q and vel}
\end{subequations}
which are shown in \cref{fig: inner velocity and Q}.

\subsection{Outer expansion}
\label{sec:outer}
The outer region corresponds to $\ep=0$, or when the disk touches the wall. To solve for $\Qt$ and $\hat{\vel}$ we will apply conformal mapping, defining the variable $\zeta=\mu+\ci \xi=-2/z$, with $\mu,\xi$ real \cite{koberDictionaryConformalRepresentations1957}. Under this map the region outside the disk transforms to the channel $\mu\in (-\infty,\infty)$, $\xi\in[0,1]$, where the plane wall maps to $\xi=0$, the disk to $\xi=1$, and the point at infinity to $\mu=\xi=0$. 

\begin{figure}[t]
\centering
\subfloat{
  \includegraphics[width=.7\linewidth]{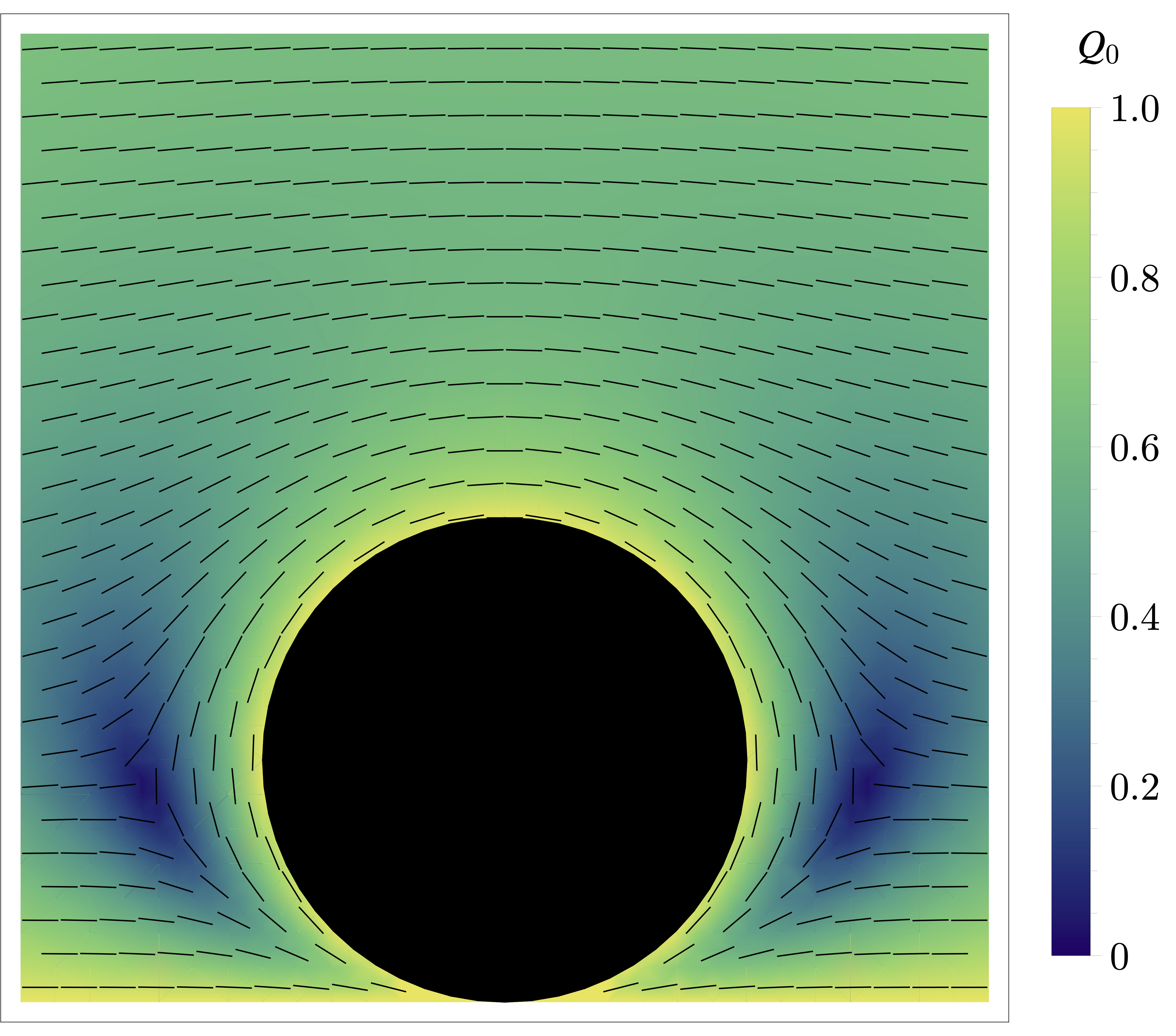}}

\subfloat{
  \centering
  \includegraphics[width=.65\linewidth]{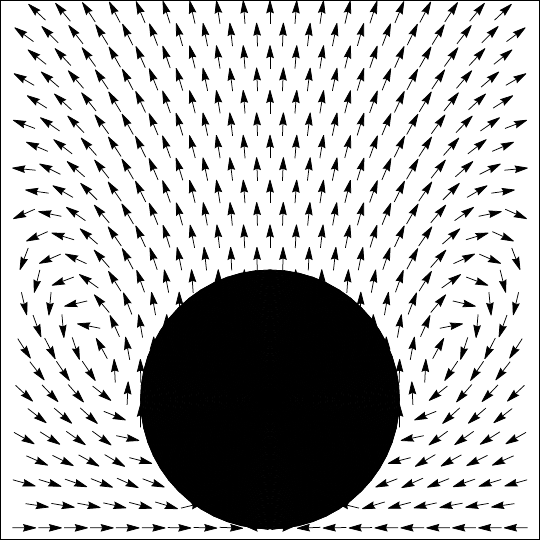}
}
\caption{Nematic director and secondary velocity in the outer limit. (Top) The nematic director is parallel anchored to all surfaces, becoming uniformly horizontal at infinity. (Bottom) The velocity diverges at the point where the disk touches the wall and so we only plot its direction.}
\label{fig: outer velocity and Q}
\end{figure}

We begin with $\Qt$, for which the boundary conditions (\ref{eqn:Q boundary conditions}) translate to 
\begin{subequations}
    \begin{gather}
        Q= 1\ \text{at}\ \xi=0,\\
        Q= \left(\frac{\mu-\ci}{\mu+\ci}\right)^2\ \text{at}\ \xi=1,
    \end{gather}
\end{subequations}
and we have used $Q=\text{Q}_1+\ci \text{Q}_2$. We solve Laplace's equation by Fourier transform in the $\mu$ direction, with the solution
\begin{equation}
    Q=1+4\int^{\infty}_{-\infty}\frac{\wrt k}{2\pi} e^{\ci k\mu-k}(k-1)\frac{\sinh k\xi}{\sinh k}.
\end{equation}
This can be integrated \cite{gradshteyn2014table} to give
\begin{equation}
\begin{aligned}
       Q&=1-2\psi(1-\ci\zeta/2)+2\psi(1-\ci\bar{\zeta}/2)\\
       &-\psi^{(1)}(1-\ci\zeta/2)+\psi^{(1)}(1-\ci\bar{\zeta}/2),
\end{aligned}
\end{equation}
where $\psi$ is the digamma function, and $\psi^{(1)}$ its derivative. This solution, shown in \cref{fig: outer velocity and Q}(a), exhibits two $-1/2$ topological defects slightly below the equator at positions $(x_{1/2},y_{1/2})\approx(\pm 1.484 , 0.876)R$. This can be compared to a disk in free space where the defects are equatorial and their position is known to be $x_{1/2}\approx 1.236R$ \cite{tasinkevych2002colloidal}.

With $\Qt$ in hand we turn to the outer velocity of the secondary problem, $\hat{\vel}$, which solves the ordinary Stokes equations with boundary conditions:
\begin{subequations}
    \begin{gather}
        \hat{\text{u}}_x+\ci \hat{\text{u}}_y=0\ \text{on the wall},\\
        \hat{\text{u}}_x+\ci \hat{\text{u}}_y=\ci \ \text{on the disk},\\
        \hat{\text{u}}_x+\ci \hat{\text{u}}_y=0\ \text{at infinity}.
    \end{gather}
\end{subequations}
There is an inconsistency in the boundary conditions where the disk touches the wall, so we search for solutions that are valid everywhere except this point. As far as we are aware the solution to this problem does not yet exist in the literature.

We begin by writing the stream function as combination of two analytic functions \cite{richardson1968two}
\begin{equation}
    \hat{\psi}=\imag[\zb f(z)+g(z)],
\end{equation}
from which the complex velocity is given by
\begin{equation}
    \hat{\text{u}}_x-\ci \hat{\text{u}}_y=\zb f'(z)+g'(z)-\overline{f(z)},
\end{equation}
where $'$ denotes a $z$ derivative. Defining $g'(z)=G(z)$ and converting to $\zeta$ coordinates yields
\begin{equation}
    \hat{\text{u}}_x-\ci \hat{\text{u}}_y = -\frac{\zeta^2}{\overline{\zeta}}f'(\zeta)+G(\zeta)-\overline{f(\zeta)},
\end{equation}
where $f(\zeta)=f(\zeta(z))$. On the disk $\overline{\zeta}=\zeta$, and on the wall $\overline{\zeta}=\zeta-2\ci$, therefore the boundary conditions may be re-written as two functional relations
\begin{subequations}
    \begin{gather}
        -\zeta f'(\zeta)+G(\zeta)-\overline{f}(\zeta)=0,\\
        \frac{\zeta^2}{2\ci-\zeta}f'(\zeta)+G(\zeta)-\overline{f}(\zeta-2\ci)=-\ci,
    \end{gather}
    \label{eqn: outer vel analytic relations}
\end{subequations}
where $\overline{f}(\zeta)=\overline{f(\overline{\zeta})}$. These two equations only depend on analytic functions of $\zeta$, and therefore hold over the entire channel. Subtracting one from the other we arrive at a functional relation for $f(\zeta)$
\begin{equation}
    2\ci\zeta f'(\zeta)+(2\ci-\zeta)[\overline{f}(\zeta)-\overline{f}(\zeta-2\ci)+\ci]=0.
\end{equation}
This is solved by the polynomial
\begin{equation}
    f(\zeta)=\frac{3\ci}{4}\zeta^2-\frac{1}{4}\zeta^3,
\label{eqn:inner f function}
\end{equation}
and (\ref{eqn:inner f function}) into  (\ref{eqn: outer vel analytic relations}) determines $G(\zeta)$ to be
\begin{equation}
    G(\zeta)=\frac{3\ci}{4}\zeta ^2-\zeta^3.
\end{equation}
With $f(\zeta)$ and $G(\zeta)$ we have the stream function and hence the full secondary velocity field. This solution, shown in Fig. \ref{fig: outer velocity and Q}(b), satisfies all the boundary conditions and although regular in the $\zeta$ plane, diverges at the touching point when converting back to real coordinates.

To check that our inner and outer solutions match we take the outer limit of the inner expansion and compare it to the inner limit of the outer expansion \cite{Hinch_1991}. We find they match, with the overlapping function given by
\begin{subequations}
\begin{gather}
    \text{Q}_{1,\text{match}}\sim  1,\\
    \text{Q}_{2,\text{match}}\sim  \frac{4y}{x},\\
    \velt_{x,\text{match}}\sim -\frac{24y}{x^3},\\
    \velt_{y,\text{match}}\sim -\frac{36 y^2}{x^4}.
\end{gather}
\end{subequations}

\subsection{Active force}
The active force is given by the integral of $\text{f}^{\: \alpha}=-\alpha\text{Q}_{ij}\nabla_i\hat{\velt}_j$ over the entire domain outside the disk. To see why we need asymptotic matching let us calculate the inner contribution to this force, which to leading order is given by 
\begin{equation}
\begin{aligned}
        \text{F}^{\alpha}_{\text{in}}=&-\alpha\ep^{1/2}\int_{-L}^{L}\int_0^{H} (\pd_X \hat{u}_0-\pd_Y\hat{v}_0)q_{1,0}\\
        &+(\pd_Y\hat{u}_0)q_{2,0}\ \wrt X\wrt Y +O(\ep),
\end{aligned}
\label{eqn: f active inner}
\end{equation}
where $L$ is some length at which the inner solution fails. In many problems this integral converges to a finite value as $L$ tends to infinity and the outer problem can be ignored \cite{cooley1969slow, jeffreySLOWMOTIONCYLINDER1981}. In our case the integral diverges, implying that we must match to an outer region
\begin{equation}
    \text{F}^{\alpha}_{\text{in}} = \alpha\ep^{1/2}[4\pi\sqrt{2}-4L]+O(L^{-1}).
\end{equation}
To perform matching we construct a globally valid asymptotic approximation to $\text{f}^{\: \alpha}$ which is then integrated over the entire domain. As $\text{f}^{\: \alpha}$ combines two functions it mixes various powers of $\ep$, which, although not a problem in the outer region ($\ep=0$), complicates matters in the inner region. To leading order $\text{f}^{\: \alpha}_{\text{inner}}$ is given by \cref{eqn: f active inner} converted back to outer, $(x,y)$ coordinates. 

To match $\text{f}^{\: \alpha}_{\text{inner}}$ to the outer integrand, $\text{f}^{\: \alpha}_{\text{outer}}$, we take the outer limit of the inner expansion and the inner limit of the outer expansion \cite{Hinch_1991}. Comparing the two we find an overlapping contribution
\begin{equation}
    \text{f}^{\alpha}_{\text{overlap}}= -\frac{48\alpha y}{x^4}+\frac{96\alpha y^2}{x^6},
\end{equation}
which upon integration over the inner region gives $\text{F}^\alpha_{\text{overlap}}\sim -4\alpha\ep^{1/2}L +O(L^{-3})$, i.e. the overlap is responsible for the divergence. This term will appear in both the inner and outer contributions to the active force but with a relative minus sign between them, cancelling when the two contributions are added.

We now construct globally valid approximation according to
\begin{equation}
\text{f}^{\alpha}_{\text{uniform}}=\text{f}^{\: \alpha}_{\text{in}}+\text{f}^{\: \alpha}_{\text{out}}-\text{f}^{\: \alpha}_{\text{overlap}},
\end{equation}
using which the total active force is $\text{F}^{\alpha}=\int \text{f}^{\: \alpha}_{\text{uniform}}\dr$, where we integrate using the conformal coordinates defined in section \ref{sec: Q tensor sol}. The result is
\begin{equation}
    \text{F}^{\alpha}\sim 4\sqrt{2}\pi R\alpha \ep^{1/2} +O(\ep),
\end{equation}
which agrees with the exact result from \cref{eqn: active force exact}. This result indicates that the leading order contribution to the active force comes from the curvature in the nematic director in a small region under the disk.

\subsection{Elastic force}
The elastic force is calculated as a surface integral using \cref{eqn: elastic integral}. The leading order inner contribution is
\begin{equation}
    \mathrm{F}^K=\hat{\mathrm{U}}_j \int_{\text {Disk }} \sigma_{i j}^K \mathrm{n}_i \mathrm{dS} \sim -\ep^{-1/2}\int_{-L}^{L}\left(\frac{\pd q_{2,0}}{\pd Y}\right)^2\wrt X.
\end{equation}
Substituting (\ref{eqn: inner Q and vel}) gives the elastic force
\begin{equation}
    \text{F}^K\sim \frac{2 \sqrt{2} K \pi}{R} \ep^{-1/2}-\frac{16 K}{R L}+O(L^{-3}),
\end{equation}
which converges to a finite value as $L$ tends to infinity and confirms the exact result of \cref{eqn: elastic force leading order}. Again, this result shows that the leading order contribution to the elastic force is due to the small region under the disk.

\section{Location and impact of topological defects}
\label{sec: defects}

Numerical studies of disks near walls in two dimensional passive liquid crystals indicate that as the disk approaches the wall, the topological defects also move closer \cite{silvestre2004colloidal}. Although these studies were based upon minimising the full Landau de-Gennes free energy (\ref{eqn: Landau de Gennes}), for non-zero $A$, our analysis indicates that some of the qualitative behaviour, such as defects moving below the equator, may be captured by setting $A$ to zero. 

In the opposite limit to what we have considered, i.e. $A$ tending to infinity, the nematic becomes infinitely stiff, and it is known that minimising the full Landau de-Gennes free energy is equivalent to solving $\nabla^2\theta=0$, where $\theta$ is the angle of the nematic director field \cite{gennesPhysicsLiquidCrystals2013}. In this limit, topological defects correspond to singularities in the angle and must be added in by hand.

\begin{figure}[t]
\centering
\subfloat{
  \includegraphics[width=.7\linewidth]{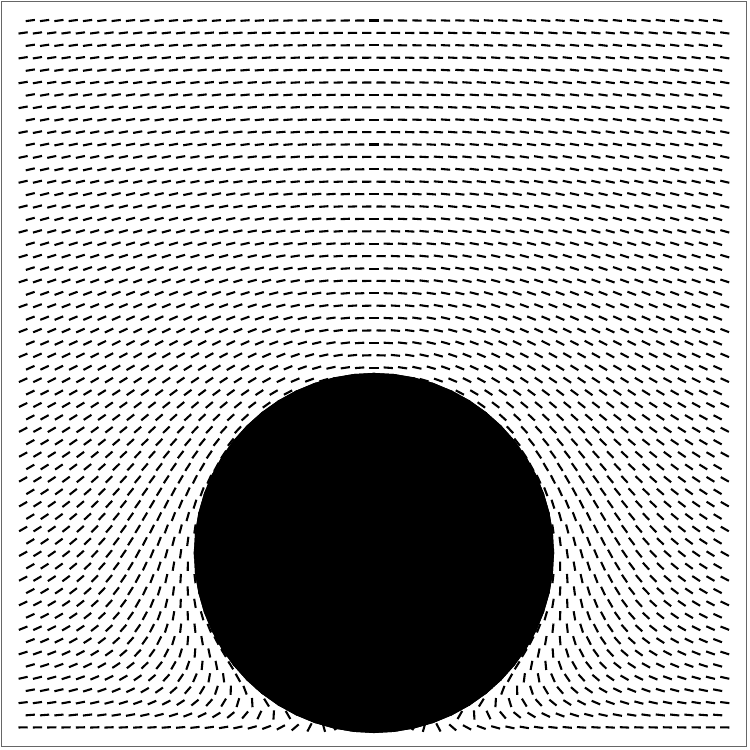}
}

\subfloat{
  \begin{tikzpicture}[scale=0.86]
\filldraw[color=blue!5,fill=blue!5] (-3.6,-3.6) rectangle (3.6,3.6);
\draw[color=blue,thick] (-3.6,-3.6)--(3.6,-3.6);
\filldraw[color=blue,fill=white, thick](0,0) circle (1.7);
\filldraw[color=blue,fill=white, thick](2.5,0) circle (.1);
\filldraw[color=blue,fill=white, thick](-2.5,0) circle (.1);
\draw[-stealth, color=blue, line width=.5mm] (0,-1.7)--(0,-2.2);
\node at (-2.6,0.3){$-1/2$};
\node at (2.4,0.3){$-1/2$};
\filldraw[color=blue,fill=white, thick](0,-3.3) circle (.1);
\node at (0,-3){$-1$};
\draw[-to,thick] (2.5,-.2) arc(0:-74:3.2) ;
\draw[-to,thick] (-2.5,-.2) arc(180:254:3.2) ;
\end{tikzpicture}
}

\caption{Director angle in the outer limit. (Top) The nematic director is parallel anchored to all surfaces, becomes uniformly horizontal at infinity, and diverges where the disk touches the plane. (Bottom) Schematic illustration of what happens to the $-1/2$ topological defects as the disk approaches and then touches the wall.}
\label{fig: outer angle and defects}
\end{figure}

It is interesting to see how our analysis with $A=0$ differs. Unfortunately we do not find it possible to use $\theta$ when the disk is above the wall due to the multiple connectivity of the domain. However when the disk is touching we may apply the Fourier transform techniques of section \ref{sec:outer}. Following the same steps we used to find $Q$ gives 
\begin{equation}
    \theta=-2 \arctan \frac{\xi}{\mu}+2 \arctan \left[\operatorname{coth} \frac{\pi \mu}{2} \tan \frac{\pi \xi}{2}\right],
\end{equation}
where $\pi/2$ may be added to switch from planar to normal anchoring. Note that $\theta$  diverges at the point where the disk touches the wall. Interestingly this solution has no visible topological defects though the disk still has a net $+1$ charge, and our $\Qt$ tensor based solution had defects. As shown in \cref{fig: outer angle and defects}, this appears to be because the two $-1/2$ topological defects move under and around the disk as it approaches, merging to create a $-1$ defect at the point of contact. This only happens in the $\theta$ description because the defects are point-like objects, whereas in in the $\Qt$ tensor description the defects always have a finite extent.

\begin{comment}
One may suspect that this lack of topological defects affects the force on the disk. However, our matched asymptotic analysis indicates that the dominant contribution to the force comes from the region under the disk where there are no defects. In fact, defining an effective $\Qt$ tensor through $Q=e^{2i\theta}$ we can perform the same asymptotic matching as earlier, and the leading order active and elastic forces are identical.
\end{comment}

\section{Conclusions}
\label{sec:discuss}

We have shown that a disk immersed in active liquid crystals can be attracted or repelled by walls depending on a combination of activity and anchoring conditions. In particular, planar (normal) anchored disks are repelled (attracted) by active forces when the nematic is contractile (extensile). When the disk is sufficiently close to the wall and the flows are sufficiently slow, we can obtain exact expressions for the effective forces by combining tools from complex analysis and the Lorentz reciprocal theorem. This theorem allowed us to avoid solving the Stokes equations forced by active stresses, and instead use a solution to the Stokes equations when activity was zero. We then applied a matched asymptotic analysis to study the same problem when the disk was very near the wall which revealed that the leading order contribution to the force came from the small region between the disk and the wall.

Our calculations are done in the limit when the active nematic affects the fluid velocity field but in which the flow does not affect the nematic director field. Nevertheless we expect qualitatively similar behaviour even when  back-flow is relevant. %\tl{due to the fact that the dominant active contribution to the effective force comes from the director configurations in a small region between the disk and the wall}. 
%(must give reason why !!)
These results can be experimentally tested in two dimensional active nematic films~\cite{activenematicreview, dogiccolloidaps, sanchez2012spontaneous, hardouin2019reconfigurable}.  
However, typically in these experiments~\cite{activenematicreview, dogiccolloidaps, sanchez2012spontaneous, hardouin2019reconfigurable} the two dimensional active nematic film lies between two other fluids and to be accurately modelled, one must include the fluid dissipation from the three-dimensional fluids above and below the film, a problem that was first considered by Saffman and Delbr\"uck~\cite{Saffman1976}. It is well known that this implies that there is two-dimensional Stokes behaviour only on length-scales much smaller than the Saffman-Delbr\"uck length~\cite{Saffman1976,Martinez-Prat2021,Matas-Navarro2014a,Levine2004,Lubensky1996,Ajdari1998}.  Hence our calculations are valid only when this length is larger than all the length-scales in the problem. Including the effect of a finite Saffman-Delbr\"uck length makes it hard to use complex analytic techniques to solve for the velocity field and one will only be able to obtain approximate results using the matched asymptotic techniques outlined above. 
We leave that for future work.
 
We foresee two further extensions of this work: the first being the problem of two disks meeting each other, rather than one approaching a wall; the second being the same problem in three dimensions, where complex analytic techniques could not be used. For the first problem, the secondary velocity field is known \cite{wakiyaApplicationBipolarCoordinates1975}, but it is difficult to solve for $\Qt$. This is a known problem, as even solving Laplace's equation for a scalar field with constant boundary conditions on two disks while going to zero at infinity is famously hard \cite{darevski1958electrostatic, lekner2021electrostatics}. For the second problem of a sphere approaching a wall, even the secondary velocity is difficult as one must use bi-spherical coordinates \cite{cooley1969slow,brenner1961slow}. Moreover, the nematic director configuration is much trickier as there is not a unique director field that is planar or normal anchored everywhere to the sphere \cite{starkPhysicsColloidalDispersions2001}. 

\section*{Acknowledgements}
The authors acknowledge helpful discussions with O. Schnitzer.
LN and TBL would like to thank the Isaac Newton Institute for Mathematical Sciences, Cambridge, for support and hospitality during the programme {\em New statistical physics in living matter}, where part of this work was done. This work was supported by EPSRC grants EP/R014604/1 and EP/T031077/1.

\section*{Author Contributions}
LN,JE,TBL designed the problem, LN performed the research and LN,JE,TBL analysed the results and wrote the manuscript.

\section*{Conflicts of interest}
There are no conflicts to declare.

\section*{Appendix: Elastic stresses}
\label{sec:appen elastic stress}
Using $\Qt$, the full equations of active nematics are \cite{cates2018theories}
\begin{subequations}
\begin{gather} 
\pd_t\Qt+\vel\cdot\nabla\Qt=-\frac{1}{\gamma}\frac{\delta \mathcal{F}}{\delta \Qt},\\
\eta\nabla^2\vel-\nabla P+\alpha\nabla\cdot\Qt+\nabla\cdot\boldsymbol{\sigma}^K=0.
\end{gather}
\label{eqn:Q eqns full}
\end{subequations}
We neglect co-rotation and flow aligning terms as their stress tensors are zero if $\nabla^2\Qt=0$.. We work with a simplified free energy
\begin{equation}
\mathcal{F}=\int\dr\frac{K}{2}(\nabla_iQ_{jk})(\nabla_iQ_{jk}),
\end{equation}
but the same method holds for more complicated free energies. To derive the elastic stress we follow \cite{cates2018theories} and calculate the rate of change of free energy.
\begin{equation}
\frac{\wrt \mathcal{F}}{\wrt t}=\int\dr\ \frac{\delta \mathcal{F}}{\delta \Qt}\pd_t\Qt=-\int\dr\ K\nabla^2\Qt\cdot \pd_t\Qt,
\end{equation}
where we have integrated by parts and dropped a surface term, assuming that $\Qt$ has fixed boundary conditions on any surface. We now replace $\pd_t\Qt$ using (\ref{eqn:Q eqns full})
\begin{equation}
\frac{\wrt \mathcal{F}}{\wrt t}=\int\dr\ \velt_j\nabla_j \Qt\cdot \nabla^2\Qt-\frac{K^2}{\gamma}|\nabla^2\Qt|^2.
\end{equation}
Now consider the total change for a small time $\delta F=\frac{\wrt F}{\wrt t}\delta t$, and balance the change in free energy due to the flow with the one caused by stress \cite{chaikin1995principles}.
\begin{equation}
\int\dr\ K\velt_j\nabla_j \Qt\cdot \nabla^2\Qt=\int\dr\ \sigma_{ij}\nabla_i \velt_j.
\end{equation}
Comparing each term we find a total stress
\begin{equation}
\boldsymbol{\sigma}^K_{ij}=-K[\nabla_i \text{Q}_{kl}\nabla_j \text{Q}_{kl}-\frac{1}{2}\delta_{ij}|\nabla_j\text{Q}_{kl}|^2].
\end{equation}

\bibliographystyle{apsrev4-1} % Tell bibtex which bibliography style to use
\bibliography{biblio}

\begin{thebibliography}{77}%
\makeatletter
\providecommand \@ifxundefined [1]{%
 \@ifx{#1\undefined}
}%
\providecommand \@ifnum [1]{%
 \ifnum #1\expandafter \@firstoftwo
 \else \expandafter \@secondoftwo
 \fi
}%
\providecommand \@ifx [1]{%
 \ifx #1\expandafter \@firstoftwo
 \else \expandafter \@secondoftwo
 \fi
}%
\providecommand \natexlab [1]{#1}%
\providecommand \enquote  [1]{``#1''}%
\providecommand \bibnamefont  [1]{#1}%
\providecommand \bibfnamefont [1]{#1}%
\providecommand \citenamefont [1]{#1}%
\providecommand \href@noop [0]{\@secondoftwo}%
\providecommand \href [0]{\begingroup \@sanitize@url \@href}%
\providecommand \@href[1]{\@@startlink{#1}\@@href}%
\providecommand \@@href[1]{\endgroup#1\@@endlink}%
\providecommand \@sanitize@url [0]{\catcode `\\12\catcode `\$12\catcode `\&12\catcode `\#12\catcode `\^12\catcode `\_12\catcode `\%12\relax}%
\providecommand \@@startlink[1]{}%
\providecommand \@@endlink[0]{}%
\providecommand \url  [0]{\begingroup\@sanitize@url \@url }%
\providecommand \@url [1]{\endgroup\@href {#1}{\urlprefix }}%
\providecommand \urlprefix  [0]{URL }%
\providecommand \Eprint [0]{\href }%
\providecommand \doibase [0]{http://dx.doi.org/}%
\providecommand \selectlanguage [0]{\@gobble}%
\providecommand \bibinfo  [0]{\@secondoftwo}%
\providecommand \bibfield  [0]{\@secondoftwo}%
\providecommand \translation [1]{[#1]}%
\providecommand \BibitemOpen [0]{}%
\providecommand \bibitemStop [0]{}%
\providecommand \bibitemNoStop [0]{.\EOS\space}%
\providecommand \EOS [0]{\spacefactor3000\relax}%
\providecommand \BibitemShut  [1]{\csname bibitem#1\endcsname}%
\let\auto@bib@innerbib\@empty
%</preamble>
\bibitem [{\citenamefont {Marchetti}\ \emph {et~al.}(2013)\citenamefont {Marchetti}, \citenamefont {Joanny}, \citenamefont {Ramaswamy}, \citenamefont {Liverpool}, \citenamefont {Prost}, \citenamefont {Rao},\ and\ \citenamefont {Simha}}]{marchettiHydrodynamicsSoftActive2013}%
  \BibitemOpen
  \bibfield  {author} {\bibinfo {author} {\bibfnamefont {M.~C.}\ \bibnamefont {Marchetti}}, \bibinfo {author} {\bibfnamefont {J.~F.}\ \bibnamefont {Joanny}}, \bibinfo {author} {\bibfnamefont {S.}~\bibnamefont {Ramaswamy}}, \bibinfo {author} {\bibfnamefont {T.~B.}\ \bibnamefont {Liverpool}}, \bibinfo {author} {\bibfnamefont {J.}~\bibnamefont {Prost}}, \bibinfo {author} {\bibfnamefont {M.}~\bibnamefont {Rao}}, \ and\ \bibinfo {author} {\bibfnamefont {R.~A.}\ \bibnamefont {Simha}},\ }\href {\doibase 10.1103/RevModPhys.85.1143} {\bibfield  {journal} {\bibinfo  {journal} {Reviews of Modern Physics}\ }\textbf {\bibinfo {volume} {85}},\ \bibinfo {pages} {1143} (\bibinfo {year} {2013})}\BibitemShut {NoStop}%
\bibitem [{\citenamefont {Ramaswamy}(2010)}]{ramaswamy2010mechanics}%
  \BibitemOpen
  \bibfield  {author} {\bibinfo {author} {\bibfnamefont {S.}~\bibnamefont {Ramaswamy}},\ }\href@noop {} {\bibfield  {journal} {\bibinfo  {journal} {Annu. Rev. Condens. Matter Phys.}\ }\textbf {\bibinfo {volume} {1}},\ \bibinfo {pages} {323} (\bibinfo {year} {2010})}\BibitemShut {NoStop}%
\bibitem [{\citenamefont {Vicsek}\ \emph {et~al.}(1995)\citenamefont {Vicsek}, \citenamefont {Czir{\'{o}}k}, \citenamefont {Ben-Jacob}, \citenamefont {Cohen},\ and\ \citenamefont {Shochet}}]{Vicsek1995}%
  \BibitemOpen
  \bibfield  {author} {\bibinfo {author} {\bibfnamefont {T.}~\bibnamefont {Vicsek}}, \bibinfo {author} {\bibfnamefont {A.}~\bibnamefont {Czir{\'{o}}k}}, \bibinfo {author} {\bibfnamefont {E.}~\bibnamefont {Ben-Jacob}}, \bibinfo {author} {\bibfnamefont {I.}~\bibnamefont {Cohen}}, \ and\ \bibinfo {author} {\bibfnamefont {O.}~\bibnamefont {Shochet}},\ }\href {http://link.aps.org/doi/10.1103/PhysRevLett.75.1226} {\bibfield  {journal} {\bibinfo  {journal} {Phys. Rev. Lett.}\ }\textbf {\bibinfo {volume} {75}},\ \bibinfo {pages} {1226} (\bibinfo {year} {1995})}\BibitemShut {NoStop}%
\bibitem [{\citenamefont {Toner}\ and\ \citenamefont {Tu}(1998)}]{toner1998flocks}%
  \BibitemOpen
  \bibfield  {author} {\bibinfo {author} {\bibfnamefont {J.}~\bibnamefont {Toner}}\ and\ \bibinfo {author} {\bibfnamefont {Y.}~\bibnamefont {Tu}},\ }\href@noop {} {\bibfield  {journal} {\bibinfo  {journal} {Physical review E}\ }\textbf {\bibinfo {volume} {58}},\ \bibinfo {pages} {4828} (\bibinfo {year} {1998})}\BibitemShut {NoStop}%
\bibitem [{\citenamefont {Toner}\ \emph {et~al.}(2005)\citenamefont {Toner}, \citenamefont {Tu},\ and\ \citenamefont {Ramaswamy}}]{toner2005hydrodynamics}%
  \BibitemOpen
  \bibfield  {author} {\bibinfo {author} {\bibfnamefont {J.}~\bibnamefont {Toner}}, \bibinfo {author} {\bibfnamefont {Y.}~\bibnamefont {Tu}}, \ and\ \bibinfo {author} {\bibfnamefont {S.}~\bibnamefont {Ramaswamy}},\ }\href@noop {} {\bibfield  {journal} {\bibinfo  {journal} {Annals of Physics}\ }\textbf {\bibinfo {volume} {318}},\ \bibinfo {pages} {170} (\bibinfo {year} {2005})}\BibitemShut {NoStop}%
\bibitem [{\citenamefont {Fily}\ and\ \citenamefont {Marchetti}(2012)}]{Fily2012}%
  \BibitemOpen
  \bibfield  {author} {\bibinfo {author} {\bibfnamefont {Y.}~\bibnamefont {Fily}}\ and\ \bibinfo {author} {\bibfnamefont {M.~C.}\ \bibnamefont {Marchetti}},\ }\href {\doibase 10.1103/PhysRevLett.108.235702} {\bibfield  {journal} {\bibinfo  {journal} {Phys. Rev. Lett.}\ }\textbf {\bibinfo {volume} {108}},\ \bibinfo {pages} {235702} (\bibinfo {year} {2012})},\ \Eprint {http://arxiv.org/abs/arXiv:1201.4847v2} {arXiv:arXiv:1201.4847v2} \BibitemShut {NoStop}%
\bibitem [{\citenamefont {Cates}\ and\ \citenamefont {Tailleur}(2015)}]{Cates2015}%
  \BibitemOpen
  \bibfield  {author} {\bibinfo {author} {\bibfnamefont {M.~E.}\ \bibnamefont {Cates}}\ and\ \bibinfo {author} {\bibfnamefont {J.}~\bibnamefont {Tailleur}},\ }\href {\doibase 10.1146/annurev-conmatphys-031214-014710} {\bibfield  {journal} {\bibinfo  {journal} {Annu. Rev. Condens. Matter Phys.}\ }\textbf {\bibinfo {volume} {6}},\ \bibinfo {pages} {219} (\bibinfo {year} {2015})},\ \Eprint {http://arxiv.org/abs/1406.3533} {arXiv:1406.3533} \BibitemShut {NoStop}%
\bibitem [{\citenamefont {Alert}\ \emph {et~al.}(2020)\citenamefont {Alert}, \citenamefont {Joanny},\ and\ \citenamefont {Casademunt}}]{alert2020universal}%
  \BibitemOpen
  \bibfield  {author} {\bibinfo {author} {\bibfnamefont {R.}~\bibnamefont {Alert}}, \bibinfo {author} {\bibfnamefont {J.-F.}\ \bibnamefont {Joanny}}, \ and\ \bibinfo {author} {\bibfnamefont {J.}~\bibnamefont {Casademunt}},\ }\href@noop {} {\bibfield  {journal} {\bibinfo  {journal} {Nature Physics}\ }\textbf {\bibinfo {volume} {16}},\ \bibinfo {pages} {682} (\bibinfo {year} {2020})}\BibitemShut {NoStop}%
\bibitem [{\citenamefont {Alert}\ \emph {et~al.}(2022)\citenamefont {Alert}, \citenamefont {Casademunt},\ and\ \citenamefont {Joanny}}]{alert2022active}%
  \BibitemOpen
  \bibfield  {author} {\bibinfo {author} {\bibfnamefont {R.}~\bibnamefont {Alert}}, \bibinfo {author} {\bibfnamefont {J.}~\bibnamefont {Casademunt}}, \ and\ \bibinfo {author} {\bibfnamefont {J.-F.}\ \bibnamefont {Joanny}},\ }\href@noop {} {\bibfield  {journal} {\bibinfo  {journal} {Annual Review of Condensed Matter Physics}\ }\textbf {\bibinfo {volume} {13}},\ \bibinfo {pages} {143} (\bibinfo {year} {2022})}\BibitemShut {NoStop}%
\bibitem [{\citenamefont {Creppy}\ \emph {et~al.}(2015)\citenamefont {Creppy}, \citenamefont {Praud}, \citenamefont {Druart}, \citenamefont {Kohnke},\ and\ \citenamefont {Plourabou{\'e}}}]{creppy2015turbulence}%
  \BibitemOpen
  \bibfield  {author} {\bibinfo {author} {\bibfnamefont {A.}~\bibnamefont {Creppy}}, \bibinfo {author} {\bibfnamefont {O.}~\bibnamefont {Praud}}, \bibinfo {author} {\bibfnamefont {X.}~\bibnamefont {Druart}}, \bibinfo {author} {\bibfnamefont {P.~L.}\ \bibnamefont {Kohnke}}, \ and\ \bibinfo {author} {\bibfnamefont {F.}~\bibnamefont {Plourabou{\'e}}},\ }\href@noop {} {\bibfield  {journal} {\bibinfo  {journal} {Physical Review E}\ }\textbf {\bibinfo {volume} {92}},\ \bibinfo {pages} {032722} (\bibinfo {year} {2015})}\BibitemShut {NoStop}%
\bibitem [{\citenamefont {Dunkel}\ \emph {et~al.}(2013)\citenamefont {Dunkel}, \citenamefont {Heidenreich}, \citenamefont {Drescher}, \citenamefont {Wensink}, \citenamefont {B{\"a}r},\ and\ \citenamefont {Goldstein}}]{dunkel2013fluid}%
  \BibitemOpen
  \bibfield  {author} {\bibinfo {author} {\bibfnamefont {J.}~\bibnamefont {Dunkel}}, \bibinfo {author} {\bibfnamefont {S.}~\bibnamefont {Heidenreich}}, \bibinfo {author} {\bibfnamefont {K.}~\bibnamefont {Drescher}}, \bibinfo {author} {\bibfnamefont {H.~H.}\ \bibnamefont {Wensink}}, \bibinfo {author} {\bibfnamefont {M.}~\bibnamefont {B{\"a}r}}, \ and\ \bibinfo {author} {\bibfnamefont {R.~E.}\ \bibnamefont {Goldstein}},\ }\href@noop {} {\bibfield  {journal} {\bibinfo  {journal} {Physical review letters}\ }\textbf {\bibinfo {volume} {110}},\ \bibinfo {pages} {228102} (\bibinfo {year} {2013})}\BibitemShut {NoStop}%
\bibitem [{\citenamefont {Gompper}\ \emph {et~al.}(2020)\citenamefont {Gompper}, \citenamefont {Winkler}, \citenamefont {Speck}, \citenamefont {Solon}, \citenamefont {Nardini}, \citenamefont {Peruani}, \citenamefont {L{\"o}wen}, \citenamefont {Golestanian}, \citenamefont {Kaupp}, \citenamefont {Alvarez} \emph {et~al.}}]{gompper20202020}%
  \BibitemOpen
  \bibfield  {author} {\bibinfo {author} {\bibfnamefont {G.}~\bibnamefont {Gompper}}, \bibinfo {author} {\bibfnamefont {R.~G.}\ \bibnamefont {Winkler}}, \bibinfo {author} {\bibfnamefont {T.}~\bibnamefont {Speck}}, \bibinfo {author} {\bibfnamefont {A.}~\bibnamefont {Solon}}, \bibinfo {author} {\bibfnamefont {C.}~\bibnamefont {Nardini}}, \bibinfo {author} {\bibfnamefont {F.}~\bibnamefont {Peruani}}, \bibinfo {author} {\bibfnamefont {H.}~\bibnamefont {L{\"o}wen}}, \bibinfo {author} {\bibfnamefont {R.}~\bibnamefont {Golestanian}}, \bibinfo {author} {\bibfnamefont {U.~B.}\ \bibnamefont {Kaupp}}, \bibinfo {author} {\bibfnamefont {L.}~\bibnamefont {Alvarez}},  \emph {et~al.},\ }\href@noop {} {\bibfield  {journal} {\bibinfo  {journal} {Journal of Physics: Condensed Matter}\ }\textbf {\bibinfo {volume} {32}},\ \bibinfo {pages} {193001} (\bibinfo {year} {2020})}\BibitemShut {NoStop}%
\bibitem [{\citenamefont {Maggi}\ \emph {et~al.}(2017)\citenamefont {Maggi}, \citenamefont {Simmchen}, \citenamefont {Saglimbeni}, \citenamefont {Katuri}, \citenamefont {Dipalo}, \citenamefont {De~Angelis}, \citenamefont {Sanchez},\ and\ \citenamefont {Di~Leonardo}}]{maggi2017self}%
  \BibitemOpen
  \bibfield  {author} {\bibinfo {author} {\bibfnamefont {C.}~\bibnamefont {Maggi}}, \bibinfo {author} {\bibfnamefont {J.}~\bibnamefont {Simmchen}}, \bibinfo {author} {\bibfnamefont {F.}~\bibnamefont {Saglimbeni}}, \bibinfo {author} {\bibfnamefont {J.}~\bibnamefont {Katuri}}, \bibinfo {author} {\bibfnamefont {M.}~\bibnamefont {Dipalo}}, \bibinfo {author} {\bibfnamefont {F.}~\bibnamefont {De~Angelis}}, \bibinfo {author} {\bibfnamefont {S.}~\bibnamefont {Sanchez}}, \ and\ \bibinfo {author} {\bibfnamefont {R.}~\bibnamefont {Di~Leonardo}},\ }\href@noop {} {\bibfield  {journal} {\bibinfo  {journal} {arXiv preprint arXiv:1707.03630}\ } (\bibinfo {year} {2017})}\BibitemShut {NoStop}%
\bibitem [{\citenamefont {Bishop}\ \emph {et~al.}(2023)\citenamefont {Bishop}, \citenamefont {Biswal},\ and\ \citenamefont {Bharti}}]{bishop2023active}%
  \BibitemOpen
  \bibfield  {author} {\bibinfo {author} {\bibfnamefont {K.~J.}\ \bibnamefont {Bishop}}, \bibinfo {author} {\bibfnamefont {S.~L.}\ \bibnamefont {Biswal}}, \ and\ \bibinfo {author} {\bibfnamefont {B.}~\bibnamefont {Bharti}},\ }\href@noop {} {\bibfield  {journal} {\bibinfo  {journal} {Annual Review of Chemical and Biomolecular Engineering}\ }\textbf {\bibinfo {volume} {14}},\ \bibinfo {pages} {1} (\bibinfo {year} {2023})}\BibitemShut {NoStop}%
\bibitem [{\citenamefont {Goodrich}\ and\ \citenamefont {Brenner}(2017)}]{goodrich2017braid}%
  \BibitemOpen
  \bibfield  {author} {\bibinfo {author} {\bibfnamefont {C.~P.}\ \bibnamefont {Goodrich}}\ and\ \bibinfo {author} {\bibfnamefont {M.~P.}\ \bibnamefont {Brenner}},\ }\href@noop {} {\bibfield  {journal} {\bibinfo  {journal} {Proceedings of the National Academy of Sciences}\ }\textbf {\bibinfo {volume} {114}},\ \bibinfo {pages} {257} (\bibinfo {year} {2017})}\BibitemShut {NoStop}%
\bibitem [{\citenamefont {Z{\"o}ttl}\ and\ \citenamefont {Stark}(2016)}]{zottl2016emergent}%
  \BibitemOpen
  \bibfield  {author} {\bibinfo {author} {\bibfnamefont {A.}~\bibnamefont {Z{\"o}ttl}}\ and\ \bibinfo {author} {\bibfnamefont {H.}~\bibnamefont {Stark}},\ }\href@noop {} {\bibfield  {journal} {\bibinfo  {journal} {Journal of Physics: Condensed Matter}\ }\textbf {\bibinfo {volume} {28}},\ \bibinfo {pages} {253001} (\bibinfo {year} {2016})}\BibitemShut {NoStop}%
\bibitem [{\citenamefont {Z{\"o}ttl}\ and\ \citenamefont {Stark}(2023)}]{zottl2023modeling}%
  \BibitemOpen
  \bibfield  {author} {\bibinfo {author} {\bibfnamefont {A.}~\bibnamefont {Z{\"o}ttl}}\ and\ \bibinfo {author} {\bibfnamefont {H.}~\bibnamefont {Stark}},\ }\href@noop {} {\bibfield  {journal} {\bibinfo  {journal} {Annual Review of Condensed Matter Physics}\ }\textbf {\bibinfo {volume} {14}},\ \bibinfo {pages} {109} (\bibinfo {year} {2023})}\BibitemShut {NoStop}%
\bibitem [{\citenamefont {Aranson}(2013)}]{aranson2013active}%
  \BibitemOpen
  \bibfield  {author} {\bibinfo {author} {\bibfnamefont {I.~S.}\ \bibnamefont {Aranson}},\ }\href@noop {} {\bibfield  {journal} {\bibinfo  {journal} {Physics-Uspekhi}\ }\textbf {\bibinfo {volume} {56}},\ \bibinfo {pages} {79} (\bibinfo {year} {2013})}\BibitemShut {NoStop}%
\bibitem [{\citenamefont {Aubret}\ \emph {et~al.}(2021)\citenamefont {Aubret}, \citenamefont {Martinet},\ and\ \citenamefont {Palacci}}]{aubret2021metamachines}%
  \BibitemOpen
  \bibfield  {author} {\bibinfo {author} {\bibfnamefont {A.}~\bibnamefont {Aubret}}, \bibinfo {author} {\bibfnamefont {Q.}~\bibnamefont {Martinet}}, \ and\ \bibinfo {author} {\bibfnamefont {J.}~\bibnamefont {Palacci}},\ }\href@noop {} {\bibfield  {journal} {\bibinfo  {journal} {Nature communications}\ }\textbf {\bibinfo {volume} {12}},\ \bibinfo {pages} {6398} (\bibinfo {year} {2021})}\BibitemShut {NoStop}%
\bibitem [{\citenamefont {{Ray}}\ \emph {et~al.}(2022)\citenamefont {{Ray}}, \citenamefont {{Zhang}}, \citenamefont {{Atzberger}}, \citenamefont {{Marchetti}},\ and\ \citenamefont {{Dogic}}}]{dogiccolloidaps}%
  \BibitemOpen
  \bibfield  {author} {\bibinfo {author} {\bibfnamefont {S.}~\bibnamefont {{Ray}}}, \bibinfo {author} {\bibfnamefont {J.}~\bibnamefont {{Zhang}}}, \bibinfo {author} {\bibfnamefont {P.}~\bibnamefont {{Atzberger}}}, \bibinfo {author} {\bibfnamefont {C.}~\bibnamefont {{Marchetti}}}, \ and\ \bibinfo {author} {\bibfnamefont {Z.}~\bibnamefont {{Dogic}}},\ }in\ \href@noop {} {\emph {\bibinfo {booktitle} {APS March Meeting Abstracts}}},\ \bibinfo {series} {APS Meeting Abstracts}, Vol.\ \bibinfo {volume} {2022}\ (\bibinfo {year} {2022})\ p.\ \bibinfo {pages} {W20.007}\BibitemShut {NoStop}%
\bibitem [{\citenamefont {Mallory}\ \emph {et~al.}(2014)\citenamefont {Mallory}, \citenamefont {Valeriani},\ and\ \citenamefont {Cacciuto}}]{mallory2014curvature}%
  \BibitemOpen
  \bibfield  {author} {\bibinfo {author} {\bibfnamefont {S.}~\bibnamefont {Mallory}}, \bibinfo {author} {\bibfnamefont {C.}~\bibnamefont {Valeriani}}, \ and\ \bibinfo {author} {\bibfnamefont {A.}~\bibnamefont {Cacciuto}},\ }\href@noop {} {\bibfield  {journal} {\bibinfo  {journal} {Physical Review E}\ }\textbf {\bibinfo {volume} {90}},\ \bibinfo {pages} {032309} (\bibinfo {year} {2014})}\BibitemShut {NoStop}%
\bibitem [{\citenamefont {Aditi~Simha}\ and\ \citenamefont {Ramaswamy}(2002)}]{simha2002hydrodynamic}%
  \BibitemOpen
  \bibfield  {author} {\bibinfo {author} {\bibfnamefont {R.}~\bibnamefont {Aditi~Simha}}\ and\ \bibinfo {author} {\bibfnamefont {S.}~\bibnamefont {Ramaswamy}},\ }\href@noop {} {\bibfield  {journal} {\bibinfo  {journal} {Physical review letters}\ }\textbf {\bibinfo {volume} {89}},\ \bibinfo {pages} {058101} (\bibinfo {year} {2002})}\BibitemShut {NoStop}%
\bibitem [{\citenamefont {Doostmohammadi}\ \emph {et~al.}(2018)\citenamefont {Doostmohammadi}, \citenamefont {Ign{\'e}s-Mullol}, \citenamefont {Yeomans},\ and\ \citenamefont {Sagu{\'e}s}}]{activenematicreview}%
  \BibitemOpen
  \bibfield  {author} {\bibinfo {author} {\bibfnamefont {A.}~\bibnamefont {Doostmohammadi}}, \bibinfo {author} {\bibfnamefont {J.}~\bibnamefont {Ign{\'e}s-Mullol}}, \bibinfo {author} {\bibfnamefont {J.~M.}\ \bibnamefont {Yeomans}}, \ and\ \bibinfo {author} {\bibfnamefont {F.}~\bibnamefont {Sagu{\'e}s}},\ }\href@noop {} {\bibfield  {journal} {\bibinfo  {journal} {Nature communications}\ }\textbf {\bibinfo {volume} {9}},\ \bibinfo {pages} {3246} (\bibinfo {year} {2018})}\BibitemShut {NoStop}%
\bibitem [{\citenamefont {Dell’Arciprete}\ \emph {et~al.}(2018)\citenamefont {Dell’Arciprete}, \citenamefont {Blow}, \citenamefont {Brown}, \citenamefont {Farrell}, \citenamefont {Lintuvuori}, \citenamefont {McVey}, \citenamefont {Marenduzzo},\ and\ \citenamefont {Poon}}]{dell2018growing}%
  \BibitemOpen
  \bibfield  {author} {\bibinfo {author} {\bibfnamefont {D.}~\bibnamefont {Dell’Arciprete}}, \bibinfo {author} {\bibfnamefont {M.~L.}\ \bibnamefont {Blow}}, \bibinfo {author} {\bibfnamefont {A.~T.}\ \bibnamefont {Brown}}, \bibinfo {author} {\bibfnamefont {F.~D.}\ \bibnamefont {Farrell}}, \bibinfo {author} {\bibfnamefont {J.~S.}\ \bibnamefont {Lintuvuori}}, \bibinfo {author} {\bibfnamefont {A.~F.}\ \bibnamefont {McVey}}, \bibinfo {author} {\bibfnamefont {D.}~\bibnamefont {Marenduzzo}}, \ and\ \bibinfo {author} {\bibfnamefont {W.~C.}\ \bibnamefont {Poon}},\ }\href@noop {} {\bibfield  {journal} {\bibinfo  {journal} {Nature communications}\ }\textbf {\bibinfo {volume} {9}},\ \bibinfo {pages} {4190} (\bibinfo {year} {2018})}\BibitemShut {NoStop}%
\bibitem [{\citenamefont {JULICHER}\ \emph {et~al.}(2007)\citenamefont {JULICHER}, \citenamefont {KRUSE}, \citenamefont {PROST},\ and\ \citenamefont {JOANNY}}]{FrenchGermanReview07}%
  \BibitemOpen
  \bibfield  {author} {\bibinfo {author} {\bibfnamefont {F.}~\bibnamefont {JULICHER}}, \bibinfo {author} {\bibfnamefont {K.}~\bibnamefont {KRUSE}}, \bibinfo {author} {\bibfnamefont {J.}~\bibnamefont {PROST}}, \ and\ \bibinfo {author} {\bibfnamefont {J.}~\bibnamefont {JOANNY}},\ }\href {\doibase 10.1016/j.physrep.2007.02.018} {\bibfield  {journal} {\bibinfo  {journal} {Phys. Rep.}\ }\textbf {\bibinfo {volume} {449}},\ \bibinfo {pages} {3} (\bibinfo {year} {2007})}\BibitemShut {NoStop}%
\bibitem [{\citenamefont {Needleman}\ and\ \citenamefont {Dogic}(2017)}]{needleman2017active}%
  \BibitemOpen
  \bibfield  {author} {\bibinfo {author} {\bibfnamefont {D.}~\bibnamefont {Needleman}}\ and\ \bibinfo {author} {\bibfnamefont {Z.}~\bibnamefont {Dogic}},\ }\href@noop {} {\bibfield  {journal} {\bibinfo  {journal} {Nature reviews materials}\ }\textbf {\bibinfo {volume} {2}},\ \bibinfo {pages} {1} (\bibinfo {year} {2017})}\BibitemShut {NoStop}%
\bibitem [{\citenamefont {Sanchez}\ \emph {et~al.}(2012)\citenamefont {Sanchez}, \citenamefont {Chen}, \citenamefont {DeCamp}, \citenamefont {Heymann},\ and\ \citenamefont {Dogic}}]{sanchez2012spontaneous}%
  \BibitemOpen
  \bibfield  {author} {\bibinfo {author} {\bibfnamefont {T.}~\bibnamefont {Sanchez}}, \bibinfo {author} {\bibfnamefont {D.~T.}\ \bibnamefont {Chen}}, \bibinfo {author} {\bibfnamefont {S.~J.}\ \bibnamefont {DeCamp}}, \bibinfo {author} {\bibfnamefont {M.}~\bibnamefont {Heymann}}, \ and\ \bibinfo {author} {\bibfnamefont {Z.}~\bibnamefont {Dogic}},\ }\href@noop {} {\bibfield  {journal} {\bibinfo  {journal} {Nature}\ }\textbf {\bibinfo {volume} {491}},\ \bibinfo {pages} {431} (\bibinfo {year} {2012})}\BibitemShut {NoStop}%
\bibitem [{\citenamefont {Mermin}(1979)}]{mermin1979topological}%
  \BibitemOpen
  \bibfield  {author} {\bibinfo {author} {\bibfnamefont {N.~D.}\ \bibnamefont {Mermin}},\ }\href@noop {} {\bibfield  {journal} {\bibinfo  {journal} {Reviews of Modern Physics}\ }\textbf {\bibinfo {volume} {51}},\ \bibinfo {pages} {591} (\bibinfo {year} {1979})}\BibitemShut {NoStop}%
\bibitem [{\citenamefont {Stark}(2001)}]{starkPhysicsColloidalDispersions2001}%
  \BibitemOpen
  \bibfield  {author} {\bibinfo {author} {\bibfnamefont {H.}~\bibnamefont {Stark}},\ }\href {\doibase 10.1016/S0370-1573(00)00144-7} {\bibfield  {journal} {\bibinfo  {journal} {Physics Reports}\ }\textbf {\bibinfo {volume} {351}},\ \bibinfo {pages} {387} (\bibinfo {year} {2001})}\BibitemShut {NoStop}%
\bibitem [{\citenamefont {Poulin}\ \emph {et~al.}(1997)\citenamefont {Poulin}, \citenamefont {Stark}, \citenamefont {Lubensky},\ and\ \citenamefont {Weitz}}]{poulin1997droplet}%
  \BibitemOpen
  \bibfield  {author} {\bibinfo {author} {\bibfnamefont {P.}~\bibnamefont {Poulin}}, \bibinfo {author} {\bibfnamefont {H.}~\bibnamefont {Stark}}, \bibinfo {author} {\bibfnamefont {T.}~\bibnamefont {Lubensky}}, \ and\ \bibinfo {author} {\bibfnamefont {D.}~\bibnamefont {Weitz}},\ }\href@noop {} {\bibfield  {journal} {\bibinfo  {journal} {Science}\ }\textbf {\bibinfo {volume} {275}},\ \bibinfo {pages} {1770} (\bibinfo {year} {1997})}\BibitemShut {NoStop}%
\bibitem [{\citenamefont {Poulin}\ and\ \citenamefont {Weitz}(1998)}]{poulin1998inverted}%
  \BibitemOpen
  \bibfield  {author} {\bibinfo {author} {\bibfnamefont {P.}~\bibnamefont {Poulin}}\ and\ \bibinfo {author} {\bibfnamefont {D.}~\bibnamefont {Weitz}},\ }\href@noop {} {\bibfield  {journal} {\bibinfo  {journal} {Physical Review E}\ }\textbf {\bibinfo {volume} {57}},\ \bibinfo {pages} {626} (\bibinfo {year} {1998})}\BibitemShut {NoStop}%
\bibitem [{\citenamefont {Mu{\v{s}}evi{\v{c}}}(2017)}]{muvsevivc2017nematic}%
  \BibitemOpen
  \bibfield  {author} {\bibinfo {author} {\bibfnamefont {I.}~\bibnamefont {Mu{\v{s}}evi{\v{c}}}},\ }\href@noop {} {\bibfield  {journal} {\bibinfo  {journal} {Materials}\ }\textbf {\bibinfo {volume} {11}},\ \bibinfo {pages} {24} (\bibinfo {year} {2017})}\BibitemShut {NoStop}%
\bibitem [{\citenamefont {Mu{\v{s}}evi{\v{c}}}\ and\ \citenamefont {{\v{S}}karabot}(2008)}]{muvsevivc2008self}%
  \BibitemOpen
  \bibfield  {author} {\bibinfo {author} {\bibfnamefont {I.}~\bibnamefont {Mu{\v{s}}evi{\v{c}}}}\ and\ \bibinfo {author} {\bibfnamefont {M.}~\bibnamefont {{\v{S}}karabot}},\ }\href@noop {} {\bibfield  {journal} {\bibinfo  {journal} {Soft Matter}\ }\textbf {\bibinfo {volume} {4}},\ \bibinfo {pages} {195} (\bibinfo {year} {2008})}\BibitemShut {NoStop}%
\bibitem [{\citenamefont {{\v{S}}karabot}\ \emph {et~al.}(2007)\citenamefont {{\v{S}}karabot}, \citenamefont {Ravnik}, \citenamefont {{\v{Z}}umer}, \citenamefont {Tkalec}, \citenamefont {Poberaj}, \citenamefont {Babi{\v{c}}}, \citenamefont {Osterman},\ and\ \citenamefont {Mu{\v{s}}evi{\v{c}}}}]{vskarabot2007two}%
  \BibitemOpen
  \bibfield  {author} {\bibinfo {author} {\bibfnamefont {M.}~\bibnamefont {{\v{S}}karabot}}, \bibinfo {author} {\bibfnamefont {M.}~\bibnamefont {Ravnik}}, \bibinfo {author} {\bibfnamefont {S.}~\bibnamefont {{\v{Z}}umer}}, \bibinfo {author} {\bibfnamefont {U.}~\bibnamefont {Tkalec}}, \bibinfo {author} {\bibfnamefont {I.}~\bibnamefont {Poberaj}}, \bibinfo {author} {\bibfnamefont {D.}~\bibnamefont {Babi{\v{c}}}}, \bibinfo {author} {\bibfnamefont {N.}~\bibnamefont {Osterman}}, \ and\ \bibinfo {author} {\bibfnamefont {I.}~\bibnamefont {Mu{\v{s}}evi{\v{c}}}},\ }\href@noop {} {\bibfield  {journal} {\bibinfo  {journal} {Physical Review E}\ }\textbf {\bibinfo {volume} {76}},\ \bibinfo {pages} {051406} (\bibinfo {year} {2007})}\BibitemShut {NoStop}%
\bibitem [{\citenamefont {Ray}\ \emph {et~al.}(2023)\citenamefont {Ray}, \citenamefont {Zhang},\ and\ \citenamefont {Dogic}}]{ray2023rectified}%
  \BibitemOpen
  \bibfield  {author} {\bibinfo {author} {\bibfnamefont {S.}~\bibnamefont {Ray}}, \bibinfo {author} {\bibfnamefont {J.}~\bibnamefont {Zhang}}, \ and\ \bibinfo {author} {\bibfnamefont {Z.}~\bibnamefont {Dogic}},\ }\href@noop {} {\bibfield  {journal} {\bibinfo  {journal} {Physical Review Letters}\ }\textbf {\bibinfo {volume} {130}},\ \bibinfo {pages} {238301} (\bibinfo {year} {2023})}\BibitemShut {NoStop}%
\bibitem [{\citenamefont {Houston}\ and\ \citenamefont {Alexander}(2023{\natexlab{a}})}]{houston2023cog}%
  \BibitemOpen
  \bibfield  {author} {\bibinfo {author} {\bibfnamefont {A.~J.}\ \bibnamefont {Houston}}\ and\ \bibinfo {author} {\bibfnamefont {G.~P.}\ \bibnamefont {Alexander}},\ }\href@noop {} {\bibfield  {journal} {\bibinfo  {journal} {arXiv preprint arXiv:2307.05247}\ } (\bibinfo {year} {2023}{\natexlab{a}})}\BibitemShut {NoStop}%
\bibitem [{\citenamefont {Loewe}\ and\ \citenamefont {Shendruk}(2022)}]{loewePassiveJanusParticles2022}%
  \BibitemOpen
  \bibfield  {author} {\bibinfo {author} {\bibfnamefont {B.}~\bibnamefont {Loewe}}\ and\ \bibinfo {author} {\bibfnamefont {T.~N.}\ \bibnamefont {Shendruk}},\ }\href {\doibase 10.1088/1367-2630/ac3b70} {\bibfield  {journal} {\bibinfo  {journal} {New Journal of Physics}\ }\textbf {\bibinfo {volume} {24}},\ \bibinfo {pages} {012001} (\bibinfo {year} {2022})}\BibitemShut {NoStop}%
\bibitem [{\citenamefont {Yao}\ \emph {et~al.}(2022)\citenamefont {Yao}, \citenamefont {Kos}, \citenamefont {Zhang}, \citenamefont {Luo}, \citenamefont {Steager}, \citenamefont {Ravnik},\ and\ \citenamefont {Stebe}}]{yao2022topological}%
  \BibitemOpen
  \bibfield  {author} {\bibinfo {author} {\bibfnamefont {T.}~\bibnamefont {Yao}}, \bibinfo {author} {\bibfnamefont {{\v{Z}}.}~\bibnamefont {Kos}}, \bibinfo {author} {\bibfnamefont {Q.~X.}\ \bibnamefont {Zhang}}, \bibinfo {author} {\bibfnamefont {Y.}~\bibnamefont {Luo}}, \bibinfo {author} {\bibfnamefont {E.~B.}\ \bibnamefont {Steager}}, \bibinfo {author} {\bibfnamefont {M.}~\bibnamefont {Ravnik}}, \ and\ \bibinfo {author} {\bibfnamefont {K.~J.}\ \bibnamefont {Stebe}},\ }\href@noop {} {\bibfield  {journal} {\bibinfo  {journal} {Science Advances}\ }\textbf {\bibinfo {volume} {8}},\ \bibinfo {pages} {eabn8176} (\bibinfo {year} {2022})}\BibitemShut {NoStop}%
\bibitem [{\citenamefont {Silvestre}\ \emph {et~al.}(2004)\citenamefont {Silvestre}, \citenamefont {Patr{\'\i}cio}, \citenamefont {Tasinkevych}, \citenamefont {Andrienko},\ and\ \citenamefont {da~Gama}}]{silvestre2004colloidal}%
  \BibitemOpen
  \bibfield  {author} {\bibinfo {author} {\bibfnamefont {N.}~\bibnamefont {Silvestre}}, \bibinfo {author} {\bibfnamefont {P.}~\bibnamefont {Patr{\'\i}cio}}, \bibinfo {author} {\bibfnamefont {M.}~\bibnamefont {Tasinkevych}}, \bibinfo {author} {\bibfnamefont {D.}~\bibnamefont {Andrienko}}, \ and\ \bibinfo {author} {\bibfnamefont {M.~T.}\ \bibnamefont {da~Gama}},\ }\href@noop {} {\bibfield  {journal} {\bibinfo  {journal} {Journal of Physics: Condensed Matter}\ }\textbf {\bibinfo {volume} {16}},\ \bibinfo {pages} {S1921} (\bibinfo {year} {2004})}\BibitemShut {NoStop}%
\bibitem [{\citenamefont {Nehari}(2012)}]{nehari2012conformal}%
  \BibitemOpen
  \bibfield  {author} {\bibinfo {author} {\bibfnamefont {Z.}~\bibnamefont {Nehari}},\ }\href@noop {} {\emph {\bibinfo {title} {Conformal mapping}}}\ (\bibinfo  {publisher} {Courier Corporation},\ \bibinfo {year} {2012})\BibitemShut {NoStop}%
\bibitem [{\citenamefont {Kober}(1957)}]{koberDictionaryConformalRepresentations1957}%
  \BibitemOpen
  \bibfield  {author} {\bibinfo {author} {\bibfnamefont {H.}~\bibnamefont {Kober}},\ }\href@noop {} {\emph {\bibinfo {title} {Dictionary of Conformal Representations}}},\ Vol.~\bibinfo {volume} {2}\ (\bibinfo  {publisher} {{Dover New York}},\ \bibinfo {year} {1957})\BibitemShut {NoStop}%
\bibitem [{\citenamefont {Chandler}\ and\ \citenamefont {Spagnolie}(2023)}]{chandler2023exact}%
  \BibitemOpen
  \bibfield  {author} {\bibinfo {author} {\bibfnamefont {T.~G.}\ \bibnamefont {Chandler}}\ and\ \bibinfo {author} {\bibfnamefont {S.~E.}\ \bibnamefont {Spagnolie}},\ }\href@noop {} {\bibfield  {journal} {\bibinfo  {journal} {arXiv preprint arXiv:2311.17708}\ } (\bibinfo {year} {2023})}\BibitemShut {NoStop}%
\bibitem [{\citenamefont {Crowdy}(2020)}]{crowdy2020solving}%
  \BibitemOpen
  \bibfield  {author} {\bibinfo {author} {\bibfnamefont {D.}~\bibnamefont {Crowdy}},\ }\href@noop {} {\emph {\bibinfo {title} {Solving problems in multiply connected domains}}}\ (\bibinfo  {publisher} {SIAM},\ \bibinfo {year} {2020})\BibitemShut {NoStop}%
\bibitem [{\citenamefont {Batista}\ \emph {et~al.}(2015)\citenamefont {Batista}, \citenamefont {Blow},\ and\ \citenamefont {da~Gama}}]{batista2015effect}%
  \BibitemOpen
  \bibfield  {author} {\bibinfo {author} {\bibfnamefont {V.~M.}\ \bibnamefont {Batista}}, \bibinfo {author} {\bibfnamefont {M.~L.}\ \bibnamefont {Blow}}, \ and\ \bibinfo {author} {\bibfnamefont {M.~M.~T.}\ \bibnamefont {da~Gama}},\ }\href@noop {} {\bibfield  {journal} {\bibinfo  {journal} {Soft Matter}\ }\textbf {\bibinfo {volume} {11}},\ \bibinfo {pages} {4674} (\bibinfo {year} {2015})}\BibitemShut {NoStop}%
\bibitem [{\citenamefont {Shendruk}\ \emph {et~al.}(2017)\citenamefont {Shendruk}, \citenamefont {Doostmohammadi}, \citenamefont {Thijssen},\ and\ \citenamefont {Yeomans}}]{shendruk2017dancing}%
  \BibitemOpen
  \bibfield  {author} {\bibinfo {author} {\bibfnamefont {T.~N.}\ \bibnamefont {Shendruk}}, \bibinfo {author} {\bibfnamefont {A.}~\bibnamefont {Doostmohammadi}}, \bibinfo {author} {\bibfnamefont {K.}~\bibnamefont {Thijssen}}, \ and\ \bibinfo {author} {\bibfnamefont {J.~M.}\ \bibnamefont {Yeomans}},\ }\href@noop {} {\bibfield  {journal} {\bibinfo  {journal} {Soft Matter}\ }\textbf {\bibinfo {volume} {13}},\ \bibinfo {pages} {3853} (\bibinfo {year} {2017})}\BibitemShut {NoStop}%
\bibitem [{\citenamefont {Beris}\ and\ \citenamefont {Edwards}(1994)}]{beris1994thermodynamics}%
  \BibitemOpen
  \bibfield  {author} {\bibinfo {author} {\bibfnamefont {A.~N.}\ \bibnamefont {Beris}}\ and\ \bibinfo {author} {\bibfnamefont {B.~J.}\ \bibnamefont {Edwards}},\ }\href@noop {} {\emph {\bibinfo {title} {Thermodynamics of flowing systems: with internal microstructure}}},\ \bibinfo {number} {36}\ (\bibinfo  {publisher} {Oxford University Press, USA},\ \bibinfo {year} {1994})\BibitemShut {NoStop}%
\bibitem [{\citenamefont {de~Gennes}\ and\ \citenamefont {Prost}(2013)}]{gennesPhysicsLiquidCrystals2013}%
  \BibitemOpen
  \bibfield  {author} {\bibinfo {author} {\bibfnamefont {P.-G.}\ \bibnamefont {de~Gennes}}\ and\ \bibinfo {author} {\bibfnamefont {J.}~\bibnamefont {Prost}},\ }\href@noop {} {\emph {\bibinfo {title} {The Physics of Liquid Crystals}}},\ \bibinfo {edition} {2nd}\ ed.,\ \bibinfo {series} {International Series of Monographs on Physics}\ No.~\bibinfo {number} {83}\ (\bibinfo  {publisher} {{Clarendon Press}},\ \bibinfo {address} {{Oxford}},\ \bibinfo {year} {2013})\BibitemShut {NoStop}%
\bibitem [{\citenamefont {Thampi}\ \emph {et~al.}(2015)\citenamefont {Thampi}, \citenamefont {Doostmohammadi}, \citenamefont {Golestanian},\ and\ \citenamefont {Yeomans}}]{thampi2015intrinsic}%
  \BibitemOpen
  \bibfield  {author} {\bibinfo {author} {\bibfnamefont {S.~P.}\ \bibnamefont {Thampi}}, \bibinfo {author} {\bibfnamefont {A.}~\bibnamefont {Doostmohammadi}}, \bibinfo {author} {\bibfnamefont {R.}~\bibnamefont {Golestanian}}, \ and\ \bibinfo {author} {\bibfnamefont {J.~M.}\ \bibnamefont {Yeomans}},\ }\href@noop {} {\bibfield  {journal} {\bibinfo  {journal} {Europhysics Letters}\ }\textbf {\bibinfo {volume} {112}},\ \bibinfo {pages} {28004} (\bibinfo {year} {2015})}\BibitemShut {NoStop}%
\bibitem [{\citenamefont {Voituriez}\ \emph {et~al.}(2005)\citenamefont {Voituriez}, \citenamefont {Joanny},\ and\ \citenamefont {Prost}}]{voituriez2005spontaneous}%
  \BibitemOpen
  \bibfield  {author} {\bibinfo {author} {\bibfnamefont {R.}~\bibnamefont {Voituriez}}, \bibinfo {author} {\bibfnamefont {J.-F.}\ \bibnamefont {Joanny}}, \ and\ \bibinfo {author} {\bibfnamefont {J.}~\bibnamefont {Prost}},\ }\href@noop {} {\bibfield  {journal} {\bibinfo  {journal} {Europhysics Letters}\ }\textbf {\bibinfo {volume} {70}},\ \bibinfo {pages} {404} (\bibinfo {year} {2005})}\BibitemShut {NoStop}%
\bibitem [{\citenamefont {Loisy}\ \emph {et~al.}(2019)\citenamefont {Loisy}, \citenamefont {Eggers},\ and\ \citenamefont {Liverpool}}]{loisy2019tractionless}%
  \BibitemOpen
  \bibfield  {author} {\bibinfo {author} {\bibfnamefont {A.}~\bibnamefont {Loisy}}, \bibinfo {author} {\bibfnamefont {J.}~\bibnamefont {Eggers}}, \ and\ \bibinfo {author} {\bibfnamefont {T.~B.}\ \bibnamefont {Liverpool}},\ }\href@noop {} {\bibfield  {journal} {\bibinfo  {journal} {Physical review letters}\ }\textbf {\bibinfo {volume} {123}},\ \bibinfo {pages} {248006} (\bibinfo {year} {2019})}\BibitemShut {NoStop}%
\bibitem [{\citenamefont {Lauga}\ and\ \citenamefont {Powers}(2009)}]{laugaHydrodynamicsSwimmingMicroorganisms2009}%
  \BibitemOpen
  \bibfield  {author} {\bibinfo {author} {\bibfnamefont {E.}~\bibnamefont {Lauga}}\ and\ \bibinfo {author} {\bibfnamefont {T.~R.}\ \bibnamefont {Powers}},\ }\href {\doibase 10.1088/0034-4885/72/9/096601} {\bibfield  {journal} {\bibinfo  {journal} {Reports on Progress in Physics}\ }\textbf {\bibinfo {volume} {72}},\ \bibinfo {pages} {096601} (\bibinfo {year} {2009})}\BibitemShut {NoStop}%
\bibitem [{\citenamefont {Crowdy}(2011)}]{crowdyTreadmillingSwimmersNoslip2011}%
  \BibitemOpen
  \bibfield  {author} {\bibinfo {author} {\bibfnamefont {D.}~\bibnamefont {Crowdy}},\ }\href@noop {} {\bibfield  {journal} {\bibinfo  {journal} {International Journal of Non-Linear Mechanics}\ }\textbf {\bibinfo {volume} {46}},\ \bibinfo {pages} {577} (\bibinfo {year} {2011})}\BibitemShut {NoStop}%
\bibitem [{\citenamefont {Chaikin}\ \emph {et~al.}(1995)\citenamefont {Chaikin}, \citenamefont {Lubensky},\ and\ \citenamefont {Witten}}]{chaikin1995principles}%
  \BibitemOpen
  \bibfield  {author} {\bibinfo {author} {\bibfnamefont {P.~M.}\ \bibnamefont {Chaikin}}, \bibinfo {author} {\bibfnamefont {T.~C.}\ \bibnamefont {Lubensky}}, \ and\ \bibinfo {author} {\bibfnamefont {T.~A.}\ \bibnamefont {Witten}},\ }\href@noop {} {\emph {\bibinfo {title} {Principles of condensed matter physics}}},\ Vol.~\bibinfo {volume} {10}\ (\bibinfo  {publisher} {Cambridge university press Cambridge},\ \bibinfo {year} {1995})\BibitemShut {NoStop}%
\bibitem [{\citenamefont {Houston}\ and\ \citenamefont {Alexander}(2023{\natexlab{b}})}]{houston2023multi}%
  \BibitemOpen
  \bibfield  {author} {\bibinfo {author} {\bibfnamefont {A.~J.}\ \bibnamefont {Houston}}\ and\ \bibinfo {author} {\bibfnamefont {G.~P.}\ \bibnamefont {Alexander}},\ }\href@noop {} {\bibfield  {journal} {\bibinfo  {journal} {Frontiers in Physics}\ }\textbf {\bibinfo {volume} {11}},\ \bibinfo {pages} {222} (\bibinfo {year} {2023}{\natexlab{b}})}\BibitemShut {NoStop}%
\bibitem [{\citenamefont {Masoud}\ and\ \citenamefont {Stone}(2019)}]{masoudReciprocalTheoremFluid2019}%
  \BibitemOpen
  \bibfield  {author} {\bibinfo {author} {\bibfnamefont {H.}~\bibnamefont {Masoud}}\ and\ \bibinfo {author} {\bibfnamefont {H.~A.}\ \bibnamefont {Stone}},\ }\href@noop {} {\bibfield  {journal} {\bibinfo  {journal} {Journal of Fluid Mechanics}\ }\textbf {\bibinfo {volume} {879}},\ \bibinfo {pages} {P1} (\bibinfo {year} {2019})}\BibitemShut {NoStop}%
\bibitem [{\citenamefont {Kim}\ and\ \citenamefont {Karrila}(2005)}]{kimMicrohydrodynamicsPrinciplesSelected2005}%
  \BibitemOpen
  \bibfield  {author} {\bibinfo {author} {\bibfnamefont {S.}~\bibnamefont {Kim}}\ and\ \bibinfo {author} {\bibfnamefont {S.~J.}\ \bibnamefont {Karrila}},\ }\href@noop {} {\emph {\bibinfo {title} {Microhydrodynamics: Principles and Selected Applications}}}\ (\bibinfo  {publisher} {{Dover Publications}},\ \bibinfo {address} {{Mineola, N.Y}},\ \bibinfo {year} {2005})\BibitemShut {NoStop}%
\bibitem [{\citenamefont {Batchelor}(2010)}]{batchelorIntroductionFluidDynamics2010}%
  \BibitemOpen
  \bibfield  {author} {\bibinfo {author} {\bibfnamefont {G.~K.}\ \bibnamefont {Batchelor}},\ }\href@noop {} {\emph {\bibinfo {title} {An {{Introduction}} to Fluid Dynamics}}},\ \bibinfo {edition} {14th}\ ed.,\ Cambridge Mathematical Library\ (\bibinfo  {publisher} {{Cambridge Univ. Press}},\ \bibinfo {address} {{Cambridge}},\ \bibinfo {year} {2010})\BibitemShut {NoStop}%
\bibitem [{\citenamefont {Jeffrey}\ and\ \citenamefont {Onishi}(1981)}]{jeffreySLOWMOTIONCYLINDER1981}%
  \BibitemOpen
  \bibfield  {author} {\bibinfo {author} {\bibfnamefont {D.~J.}\ \bibnamefont {Jeffrey}}\ and\ \bibinfo {author} {\bibfnamefont {Y.}~\bibnamefont {Onishi}},\ }\href {\doibase 10.1093/qjmam/34.2.129} {\bibfield  {journal} {\bibinfo  {journal} {The Quarterly Journal of Mechanics and Applied Mathematics}\ }\textbf {\bibinfo {volume} {34}},\ \bibinfo {pages} {129} (\bibinfo {year} {1981})}\BibitemShut {NoStop}%
\bibitem [{\citenamefont {Cooley}\ and\ \citenamefont {O'neill}(1969)}]{cooley1969slow}%
  \BibitemOpen
  \bibfield  {author} {\bibinfo {author} {\bibfnamefont {M.}~\bibnamefont {Cooley}}\ and\ \bibinfo {author} {\bibfnamefont {M.}~\bibnamefont {O'neill}},\ }\href@noop {} {\bibfield  {journal} {\bibinfo  {journal} {Mathematika}\ }\textbf {\bibinfo {volume} {16}},\ \bibinfo {pages} {37} (\bibinfo {year} {1969})}\BibitemShut {NoStop}%
\bibitem [{\citenamefont {Bender}\ and\ \citenamefont {Orszag}(2013)}]{bender2013advanced}%
  \BibitemOpen
  \bibfield  {author} {\bibinfo {author} {\bibfnamefont {C.~M.}\ \bibnamefont {Bender}}\ and\ \bibinfo {author} {\bibfnamefont {S.~A.}\ \bibnamefont {Orszag}},\ }\href@noop {} {\emph {\bibinfo {title} {Advanced mathematical methods for scientists and engineers I: Asymptotic methods and perturbation theory}}}\ (\bibinfo  {publisher} {Springer Science \& Business Media},\ \bibinfo {year} {2013})\BibitemShut {NoStop}%
\bibitem [{\citenamefont {Hinch}(1991)}]{Hinch_1991}%
  \BibitemOpen
  \bibfield  {author} {\bibinfo {author} {\bibfnamefont {E.~J.}\ \bibnamefont {Hinch}},\ }\href@noop {} {\emph {\bibinfo {title} {Perturbation Methods}}},\ Cambridge Texts in Applied Mathematics\ (\bibinfo  {publisher} {Cambridge University Press},\ \bibinfo {year} {1991})\BibitemShut {NoStop}%
\bibitem [{\citenamefont {Joanny}\ and\ \citenamefont {Ramaswamy}(2012)}]{joanny2012drop}%
  \BibitemOpen
  \bibfield  {author} {\bibinfo {author} {\bibfnamefont {J.-F.}\ \bibnamefont {Joanny}}\ and\ \bibinfo {author} {\bibfnamefont {S.}~\bibnamefont {Ramaswamy}},\ }\href@noop {} {\bibfield  {journal} {\bibinfo  {journal} {Journal of fluid mechanics}\ }\textbf {\bibinfo {volume} {705}},\ \bibinfo {pages} {46} (\bibinfo {year} {2012})}\BibitemShut {NoStop}%
\bibitem [{\citenamefont {Gradshteyn}\ and\ \citenamefont {Ryzhik}(2014)}]{gradshteyn2014table}%
  \BibitemOpen
  \bibfield  {author} {\bibinfo {author} {\bibfnamefont {I.~S.}\ \bibnamefont {Gradshteyn}}\ and\ \bibinfo {author} {\bibfnamefont {I.~M.}\ \bibnamefont {Ryzhik}},\ }\href@noop {} {\emph {\bibinfo {title} {Table of integrals, series, and products}}}\ (\bibinfo  {publisher} {Academic press},\ \bibinfo {year} {2014})\BibitemShut {NoStop}%
\bibitem [{\citenamefont {Tasinkevych}\ \emph {et~al.}(2002)\citenamefont {Tasinkevych}, \citenamefont {Silvestre}, \citenamefont {Patricio},\ and\ \citenamefont {Telo~da Gama}}]{tasinkevych2002colloidal}%
  \BibitemOpen
  \bibfield  {author} {\bibinfo {author} {\bibfnamefont {M.}~\bibnamefont {Tasinkevych}}, \bibinfo {author} {\bibfnamefont {N.}~\bibnamefont {Silvestre}}, \bibinfo {author} {\bibfnamefont {P.}~\bibnamefont {Patricio}}, \ and\ \bibinfo {author} {\bibfnamefont {M.}~\bibnamefont {Telo~da Gama}},\ }\href@noop {} {\bibfield  {journal} {\bibinfo  {journal} {The European Physical Journal E}\ }\textbf {\bibinfo {volume} {9}},\ \bibinfo {pages} {341} (\bibinfo {year} {2002})}\BibitemShut {NoStop}%
\bibitem [{\citenamefont {Richardson}(1968)}]{richardson1968two}%
  \BibitemOpen
  \bibfield  {author} {\bibinfo {author} {\bibfnamefont {S.}~\bibnamefont {Richardson}},\ }\href@noop {} {\bibfield  {journal} {\bibinfo  {journal} {Journal of Fluid Mechanics}\ }\textbf {\bibinfo {volume} {33}},\ \bibinfo {pages} {475} (\bibinfo {year} {1968})}\BibitemShut {NoStop}%
\bibitem [{\citenamefont {Hardo{\"u}in}\ \emph {et~al.}(2019)\citenamefont {Hardo{\"u}in}, \citenamefont {Hughes}, \citenamefont {Doostmohammadi}, \citenamefont {Laurent}, \citenamefont {Lopez-Leon}, \citenamefont {Yeomans}, \citenamefont {Ign{\'e}s-Mullol},\ and\ \citenamefont {Sagu{\'e}s}}]{hardouin2019reconfigurable}%
  \BibitemOpen
  \bibfield  {author} {\bibinfo {author} {\bibfnamefont {J.}~\bibnamefont {Hardo{\"u}in}}, \bibinfo {author} {\bibfnamefont {R.}~\bibnamefont {Hughes}}, \bibinfo {author} {\bibfnamefont {A.}~\bibnamefont {Doostmohammadi}}, \bibinfo {author} {\bibfnamefont {J.}~\bibnamefont {Laurent}}, \bibinfo {author} {\bibfnamefont {T.}~\bibnamefont {Lopez-Leon}}, \bibinfo {author} {\bibfnamefont {J.~M.}\ \bibnamefont {Yeomans}}, \bibinfo {author} {\bibfnamefont {J.}~\bibnamefont {Ign{\'e}s-Mullol}}, \ and\ \bibinfo {author} {\bibfnamefont {F.}~\bibnamefont {Sagu{\'e}s}},\ }\href@noop {} {\bibfield  {journal} {\bibinfo  {journal} {Communications Physics}\ }\textbf {\bibinfo {volume} {2}},\ \bibinfo {pages} {121} (\bibinfo {year} {2019})}\BibitemShut {NoStop}%
\bibitem [{\citenamefont {Saffman}(1976)}]{Saffman1976}%
  \BibitemOpen
  \bibfield  {author} {\bibinfo {author} {\bibfnamefont {P.~G.}\ \bibnamefont {Saffman}},\ }\href {\doibase 10.1017/S0022112076001511} {\bibfield  {journal} {\bibinfo  {journal} {J. Fluid Mech.}\ }\textbf {\bibinfo {volume} {73}},\ \bibinfo {pages} {593} (\bibinfo {year} {1976})}\BibitemShut {NoStop}%
\bibitem [{\citenamefont {Mart{\'{i}}nez-Prat}\ \emph {et~al.}(2021)\citenamefont {Mart{\'{i}}nez-Prat}, \citenamefont {Alert}, \citenamefont {Meng}, \citenamefont {Ign{\'{e}}s-Mullol}, \citenamefont {Joanny}, \citenamefont {Casademunt}, \citenamefont {Golestanian},\ and\ \citenamefont {Sagu{\'{e}}s}}]{Martinez-Prat2021}%
  \BibitemOpen
  \bibfield  {author} {\bibinfo {author} {\bibfnamefont {B.}~\bibnamefont {Mart{\'{i}}nez-Prat}}, \bibinfo {author} {\bibfnamefont {R.}~\bibnamefont {Alert}}, \bibinfo {author} {\bibfnamefont {F.}~\bibnamefont {Meng}}, \bibinfo {author} {\bibfnamefont {J.}~\bibnamefont {Ign{\'{e}}s-Mullol}}, \bibinfo {author} {\bibfnamefont {J.-F.}\ \bibnamefont {Joanny}}, \bibinfo {author} {\bibfnamefont {J.}~\bibnamefont {Casademunt}}, \bibinfo {author} {\bibfnamefont {R.}~\bibnamefont {Golestanian}}, \ and\ \bibinfo {author} {\bibfnamefont {F.}~\bibnamefont {Sagu{\'{e}}s}},\ }\href {\doibase 10.1103/PhysRevX.11.031065} {\bibfield  {journal} {\bibinfo  {journal} {Phys. Rev. X}\ }\textbf {\bibinfo {volume} {11}},\ \bibinfo {pages} {031065} (\bibinfo {year} {2021})},\ \Eprint {http://arxiv.org/abs/2101.11570} {arXiv:2101.11570} \BibitemShut {NoStop}%
\bibitem [{\citenamefont {Matas-Navarro}\ \emph {et~al.}(2014)\citenamefont {Matas-Navarro}, \citenamefont {Golestanian}, \citenamefont {Liverpool},\ and\ \citenamefont {Fielding}}]{Matas-Navarro2014a}%
  \BibitemOpen
  \bibfield  {author} {\bibinfo {author} {\bibfnamefont {R.}~\bibnamefont {Matas-Navarro}}, \bibinfo {author} {\bibfnamefont {R.}~\bibnamefont {Golestanian}}, \bibinfo {author} {\bibfnamefont {T.~B.}\ \bibnamefont {Liverpool}}, \ and\ \bibinfo {author} {\bibfnamefont {S.~M.}\ \bibnamefont {Fielding}},\ }\href {\doibase 10.1103/PhysRevE.90.032304} {\bibfield  {journal} {\bibinfo  {journal} {Phys. Rev. E}\ }\textbf {\bibinfo {volume} {90}},\ \bibinfo {pages} {032304} (\bibinfo {year} {2014})}\BibitemShut {NoStop}%
\bibitem [{\citenamefont {Levine}\ \emph {et~al.}(2004)\citenamefont {Levine}, \citenamefont {Liverpool},\ and\ \citenamefont {MacKintosh}}]{Levine2004}%
  \BibitemOpen
  \bibfield  {author} {\bibinfo {author} {\bibfnamefont {A.}~\bibnamefont {Levine}}, \bibinfo {author} {\bibfnamefont {T.}~\bibnamefont {Liverpool}}, \ and\ \bibinfo {author} {\bibfnamefont {F.}~\bibnamefont {MacKintosh}},\ }\href {\doibase 10.1103/PhysRevE.69.021503} {\bibfield  {journal} {\bibinfo  {journal} {Phys. Rev. E}\ }\textbf {\bibinfo {volume} {69}},\ \bibinfo {pages} {1} (\bibinfo {year} {2004})}\BibitemShut {NoStop}%
\bibitem [{\citenamefont {Lubensky}\ and\ \citenamefont {Goldstein}(1996)}]{Lubensky1996}%
  \BibitemOpen
  \bibfield  {author} {\bibinfo {author} {\bibfnamefont {D.~K.}\ \bibnamefont {Lubensky}}\ and\ \bibinfo {author} {\bibfnamefont {R.~E.}\ \bibnamefont {Goldstein}},\ }\href {\doibase 10.1063/1.868893} {\bibfield  {journal} {\bibinfo  {journal} {Phys. Fluids}\ }\textbf {\bibinfo {volume} {8}},\ \bibinfo {pages} {843} (\bibinfo {year} {1996})},\ \Eprint {http://arxiv.org/abs/9602002} {arXiv:9602002 [patt-sol]} \BibitemShut {NoStop}%
\bibitem [{\citenamefont {STONE}\ and\ \citenamefont {AJDARI}(1998)}]{Ajdari1998}%
  \BibitemOpen
  \bibfield  {author} {\bibinfo {author} {\bibfnamefont {H.~A.}\ \bibnamefont {STONE}}\ and\ \bibinfo {author} {\bibfnamefont {A.}~\bibnamefont {AJDARI}},\ }\href {\doibase 10.1017/S0022112098001980} {\bibfield  {journal} {\bibinfo  {journal} {J. Fluid Mech.}\ }\textbf {\bibinfo {volume} {369}},\ \bibinfo {pages} {151} (\bibinfo {year} {1998})}\BibitemShut {NoStop}%
\bibitem [{\citenamefont {Wakiya}(1975)}]{wakiyaApplicationBipolarCoordinates1975}%
  \BibitemOpen
  \bibfield  {author} {\bibinfo {author} {\bibfnamefont {S.}~\bibnamefont {Wakiya}},\ }\href {\doibase 10.1143/JPSJ.39.1603} {\bibfield  {journal} {\bibinfo  {journal} {Journal of the Physical Society of Japan}\ }\textbf {\bibinfo {volume} {39}},\ \bibinfo {pages} {1603} (\bibinfo {year} {1975})}\BibitemShut {NoStop}%
\bibitem [{\citenamefont {Darevski}(1958)}]{darevski1958electrostatic}%
  \BibitemOpen
  \bibfield  {author} {\bibinfo {author} {\bibfnamefont {A.}~\bibnamefont {Darevski}},\ }\href@noop {} {\enquote {\bibinfo {title} {The electrostatic field of a split phase},}\ } (\bibinfo {year} {1958})\BibitemShut {NoStop}%
\bibitem [{\citenamefont {Lekner}(2021)}]{lekner2021electrostatics}%
  \BibitemOpen
  \bibfield  {author} {\bibinfo {author} {\bibfnamefont {J.}~\bibnamefont {Lekner}},\ }\href@noop {} {\emph {\bibinfo {title} {Electrostatics of Conducting Cylinders and Spheres}}}\ (\bibinfo  {publisher} {AIP Publishing},\ \bibinfo {year} {2021})\BibitemShut {NoStop}%
\bibitem [{\citenamefont {Brenner}(1961)}]{brenner1961slow}%
  \BibitemOpen
  \bibfield  {author} {\bibinfo {author} {\bibfnamefont {H.}~\bibnamefont {Brenner}},\ }\href@noop {} {\bibfield  {journal} {\bibinfo  {journal} {Chemical engineering science}\ }\textbf {\bibinfo {volume} {16}},\ \bibinfo {pages} {242} (\bibinfo {year} {1961})}\BibitemShut {NoStop}%
\bibitem [{\citenamefont {Cates}\ and\ \citenamefont {Tjhung}(2018)}]{cates2018theories}%
  \BibitemOpen
  \bibfield  {author} {\bibinfo {author} {\bibfnamefont {M.~E.}\ \bibnamefont {Cates}}\ and\ \bibinfo {author} {\bibfnamefont {E.}~\bibnamefont {Tjhung}},\ }\href@noop {} {\bibfield  {journal} {\bibinfo  {journal} {Journal of Fluid Mechanics}\ }\textbf {\bibinfo {volume} {836}},\ \bibinfo {pages} {P1} (\bibinfo {year} {2018})}\BibitemShut {NoStop}%
\end{thebibliography}%
\end{document}